\documentclass[twocolumn,aps,prl,superscriptaddress]{revtex4}

\usepackage{amsmath}
\usepackage{graphicx}
\usepackage{MnSymbol}

\begin{document}

\title{Influence of temperature and crack-tip speed on crack propagation in elastic solids}

\author{B.N.J. Persson}
\affiliation{State Key Laboratory of Solid Lubrication, Lanzhou Institute of Chemical Physics, Chinese Academy of Sciences, 730000 Lanzhou, China}
\affiliation{Peter Gr\"unberg Institute (PGI-1), Forschungszentrum J\"ulich, 52425, J\"ulich, Germany}
\affiliation{MultiscaleConsulting, Wolfshovener str. 2, 52428 J\"ulich, Germany}

\begin{abstract}
I study the influence of temperature and the crack-tip velocity of the bond breaking
at the crack tip in rubber-like materials. The bond breaking is considered as a stress-aided
thermally activated process and result in an effective crack propagation energy which 
increases strongly with decreasing temperature or increasing crack-tip speed.
This effect is particular important for adhesive (interfacial) crack propagation but less important
for cohesive crack propagation owing to the much larger bond-breaking energies in the latter case.
For  adhesive cracks the theory results are consistent with adhesion measurements for silicone 
(PDMS) rubber in contact with silica glass surfaces. For cohesive cracks the theory agree well with
experimental results PDMS films chemically bound to silinized glass.
\end{abstract}

\maketitle

\pagestyle{empty}


\vskip 0.3cm
{\bf 1 Introduction}

Crack propagation in rubber like materials is a topic of great importance
and is involved in many practical applications such as the wear of tires or conveyor belts\cite{Gert0,Gert00}.
The energy to propagate an opening crack over a surface area $\Delta A$ in a viscoelastic solids
depends on the temperature $T$ and the crack-tip velocity $v$ and is usually written as
$G(v,T) \Delta A$. We will for simplicity 
denote $G(v,T)$ as the ``crack-energy''. The crack-energy is usually written as\cite{Gent,many,Muser,Riv,Cristiano,Carb,Creton2,Horst,Berth}
$$G(v,T) = G_0(v,T) [1+f(v,T)]\eqno(1)$$
Here the factor $[1+f(v,T)]$ is attributed to 
viscoelastic energy dissipation in a region which can extend far from the
crack tip\cite{Knaus2,Sch,adhesion,Kramer1,Gennes,Brener,Gert,Per1,Per2,hot1,hot2,CarbL1,CarbL2,MuesL1,Cia} 
and $G_0(v,T)$ is due to the bond breaking in the crack-tip process zone. For cohesive cracks (cracks propagating inside the rubber)
recent studies\cite{Creton1} have shown that the bond breaking may start far from the crack tip and in these cases there may be no clear
separation between the process zone and the viscoelastic dissipation zone. For adhesive crack propagation at the interface between a viscoelastic
solid and a counter surface the stresses may be much smaller than for cohesive crack propagation and in these cases there may be a
negligible overlap between the two spatial regions, but I am not aware of experimental studies addressing this topic.

\begin{figure}
\centering
\includegraphics[width=0.95\columnwidth]{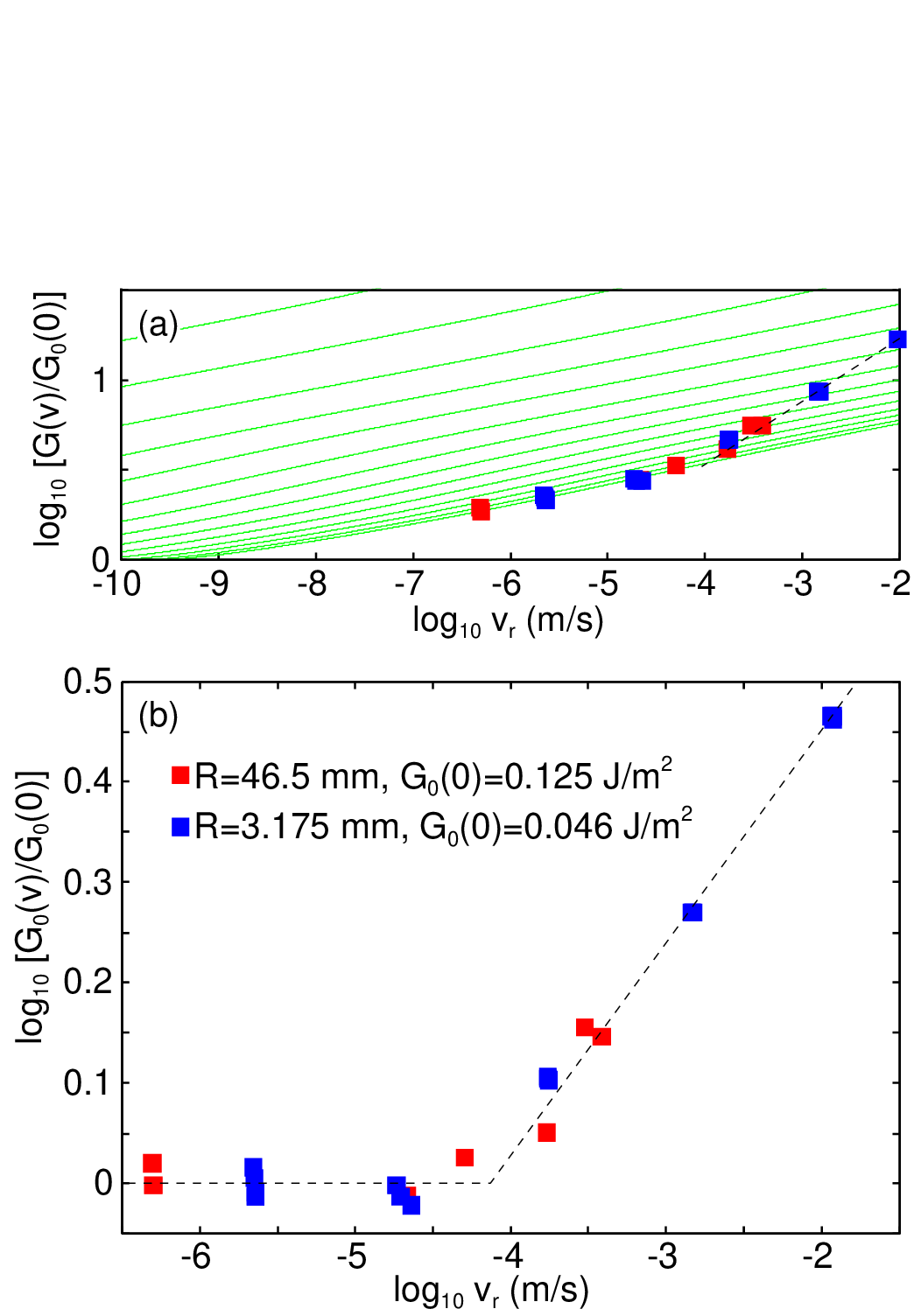}
\caption{\label{Combined.ps}
The square symbols are the measured crack-energy $G(v)=G_0(v)[1+f(v)]$ (a) 
and $G_0(v)$ (b) as a function of the crack-tip velocity $v$
(Log-Log scale) for a polydimethylsiloxane (PDMS) rubber 
(Sylgard 184, 1:10 cross-linker/base) at $T=20^\circ {\rm C}$. 
The green lines are the calculated $[1+f(v)]$ factor assuming a constant (velocity independent) $G_0(v)$.
The lowest green line is for $T=20^\circ {\rm C}$ and the other lines for $10$, $0$, .., $-110^\circ {\rm C}$.
Adapted from Ref. \cite{one}.
}
\end{figure}

In the present study I will assume that there is negligible viscoelastic energy dissipation and focus on the velocity and temperature dependency of the
$G_0(v,T)$ factor in (1). Silicone rubber has very low glass transition temperature ($T_{\rm g} \approx - 120^\circ {\rm C}$). At room
temperature, and for typical crack-tip speeds, polydimethylsiloxane (PDMS) rubber behave as a nearly perfect elastic material. 

In the following when we focuus mainly on the velocity dependency of the crack-energy 
and we denote $G(v,T)$ by $G(v)$ for simplicity.
In Fig. \ref{Combined.ps}(a) the square symbols are the measured crack-energy $G(v)=G_0(v)[1+f(v)]$ and (b)
$G_0(v)$ as a function of the crack-tip velocity $v$, for PDMS in contact with silica glass at $T=20^\circ {\rm C}$.
The green lines are the calculated $[1+f(v)]$ factor for different temperatures 
assuming a constant (velocity independent) $G_0(v)$.
The lowest green line is for $T=20^\circ {\rm C}$ and agree well with the measured data for $v<v_{\rm c} \approx 0.1 \ {\rm mm/s}$.
Similar results was obtained in Ref. \cite{Cia1,Cros}, and in
Ref. \cite{two} for PDMS with other cross-link densities. Note the abrupt increase in $G(v)$
for $v > v_{\rm c}$. We will show that for $v< v_{\rm c}$ the crack-energy $G_0(v)$ is the thermal equilibrium bond-breaking energy,
while for $v>v_{\rm c}$ the bond-breaking occur in a non-adiabatic way resulting in an increase in $G_0(v)$ with increasing $v$.

At very low crack-tip speeds bonds are broken by thermal fluctuations rather than by 
the applied force. For this case 
the breaking of adhesive bonds may require relative small energies which can be estimated
from the low-velocity (adiabatic) crack-energy $G_0(v\rightarrow 0)$ which 
for silicone rubber against a silica glass surface may be of order $\sim 0.1 \ {\rm J/m^2}$ or
$\sim 0.6 \ {\rm eV/nm^2}$. Assuming that the bonding area is of order $\sim 1 \ {\rm nm^2}$ 
this gives effective bond energies of order $\sim 0.6 \ {\rm eV}$. 
Note that this is larger than expected for 
a single Van der Waals bond; the detached surface area is 
not of the size of a single atom but most likely
a nanometer sized patch of the rubber. For cohesive cracks strong 
covalent bonds are broken at the crack tip which may require $\sim 3 \ {\rm eV}$. For this case
the effective energy to break bonds will increase with the crack-tip speed already at very
low crack-tip speeds, while for PDMS at room temperature this occur only for relative large
crack-tip speeds of order $0.1 \ {\rm mm/s}$ (see Fig. \ref{Combined.ps}).

\begin{figure}
\centering
\includegraphics[width=0.7\columnwidth]{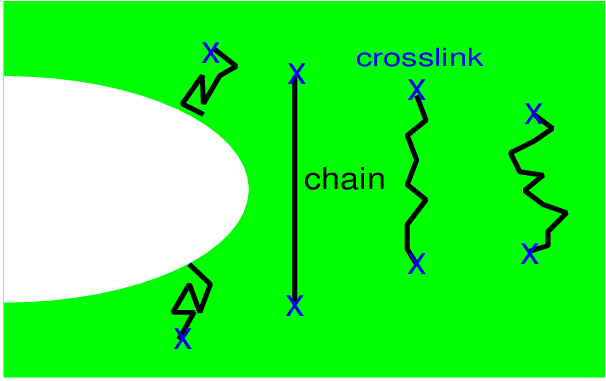}
\caption{\label{breakchain.eps}
The Lake and Thomas explanation of the high crack-energy of rubber-like materials.
When the crack-tip approach a chain molecule (consisting of $N$ monomers or bond-units 
between the crosslinks), 
extending over the (to be) fracture plane,
the chain stretch and at the point when one bond break the stored up energy is $N$
times the energy to break one bond. 
}
\end{figure}

In a fundamental study Lake and Thomas\cite{LT} have shown that for highly elastic materials
at high crack-tip speeds, where
thermal effects are less important, for cohesive cracks the crack-energy
may be strongly enhanced (by a factor up to $\sim 100$) compared to its (low-velocity)
adiabatic value (which may never be observed experimentally as it require lower velocities than
can ever be reached on the time scale of human activities). 
They showed that this enhancement is due to the stretching of polymer chains before
breaking the strong covalent bonds (see Fig. \ref{breakchain.eps}). 
This effect is also important for adhesive cracks but
in these cases only for much higher crack-tip speeds e.g. above $0.1 \ {\rm mm/s}$ for
PDMS against silica glass.

In this paper I study the influence of temperature and the crack-tip velocity of the bond breaking
at the crack tip in rubber-like materials. The bond breaking is considered as a stress-aided
thermally activated process and result in an effective crack-energy which 
increases strongly with decreasing temperature or increasing crack-tip speed.
The analysis includes the chain-stretching effect of Lake and Thomas and result in 
a $G_0(v,T)$ which is consistent with experimental observations. 
I note that stress-aided thermally activated processes are involved in nearly all applications
where bonds are broken, e.g., in sliding friction\cite{also1,also2,also3,sliding,MuesL2}, 
and goes back to the theories of
Prandtl\cite{also3} 
(see also Ref. \cite{sliding,MuesL2}) and Eyring\cite{Ey} for the dependency of the strength of 
solids (plasticity) and fluids (viscosity) on the temperature and the deformation rate. 
Other related work are theoretical (and Atomic
Force Microscopy) studies of the strength of single bonds\cite{Evans1,Evans2}.
Kendal\cite{Ke} and Chaudhury\cite{Ch,Chaud1,Chaud2} (see also \cite{Wang1,Wang2})
have suggested that stress-aided thermally activated 
bond breaking may be important also for adhesive cracks.

\vskip 0.3cm
{\bf 2 Crack propagation in a long slab--energy method}

\begin{figure}
\centering
\includegraphics[width=0.95\columnwidth]{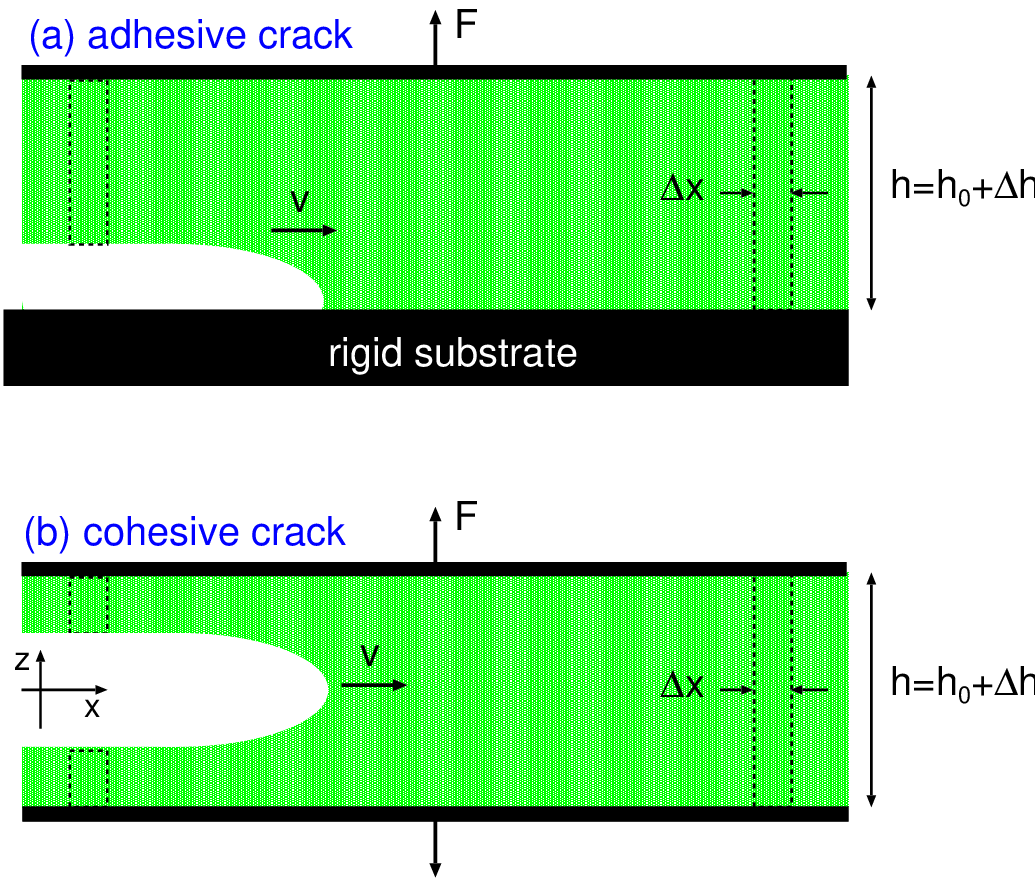}
\caption{\label{CrackBlockThree.eps}
(a) Adhesive crack: A crack at the interface between a rigid substrate and an elastic block in adhesive contact with the
substrate (adhesive crack). The rectangular elastic block has the with length $L$, height $h_0$ and width (not shown) $w$.
We assume $L>>h_0$ and $h_0 >> w$. 
The top surface of the block is clamped (glued to rigid plate, upper black region) and displaced
so the height of the block increases with $\Delta h$. The stress in the block far in front of the
crack is $\sigma_0 = E \epsilon$ where the strain $\epsilon=\Delta h/h_0$. The stress far behind the crack tip
vanish. (b) Cohesive crack: A crack occur in the middle of the elastic block.
}
\end{figure}

Consider an elastic rectangular block with the length $L$ much larger than the width $h_0$ in adhesive contact with
a substrate. We assume that the upper 
surface of the slab is clamped and displaced by $ \Delta h$
as indicated in Fig. \ref{CrackBlockThree.eps}. 

{\it Adhesive} crack propagation occur at the interface between an elastic block (or film) 
and another solid (substrate), which usually
can be considered as rigid, see Fig. \ref{CrackBlockThree.eps}(a). In this case the crack-energy
$G$ will involve breaking the adhesive bonds at the interface between the solids, 
and the elastic energy $U_{\rm el}$ is stored mainly in the upper (elastic) block. 
For {\it cohesive} crack propagation [see Fig. \ref{CrackBlockThree.eps}(b)] the upper $z>0$ and lower $z<0$ part of the solid block
are the same. 

Assume that a crack occur for 
$x<0$ and that the crack is much longer than the width $h_0$ of the slab.
A segment of the block of width $\Delta x$ for $x>>h_0$ will experience a uniform 
strain $\epsilon_0 = \Delta h/ h_0$ and the stress $\sigma_0 = E \epsilon_0$ 
so the elastic deformation energy stored up in the segment is
$$U_{\rm el} = {1\over 2} \sigma_0 \epsilon_0 \Delta V = {1\over 2 E} \sigma_0^2 \Delta V$$
where $\Delta V = h_0 w \Delta x$ is the volume of the segment ($w$ is the thickness of the slab).
Far behind the crack tip (for $x << -h_0$) the strain vanish and the elastic energy in a segment
of width $\Delta x$ vanish. If the crack moves forwards with a distance $\Delta x$ the
bonds between the atoms must be broken over an area $\Delta A = w \Delta x$. If $G$ denote the
energy to break the bonds per unit surface area then the fracture energy
$$U_{\rm ad} = G w \Delta x$$
In order for the crack to be able to move forward the reduction in the stored elastic energy must be (at least)
the energy needed to break the interfacial bonds or $U_{\rm el} = U_{\rm ad}$ giving
$$\sigma_0 = \left ( {2G E \over h_0}\right )^{1/2}\eqno(2)$$

\begin{figure}
\centering
\includegraphics[width=0.95\columnwidth]{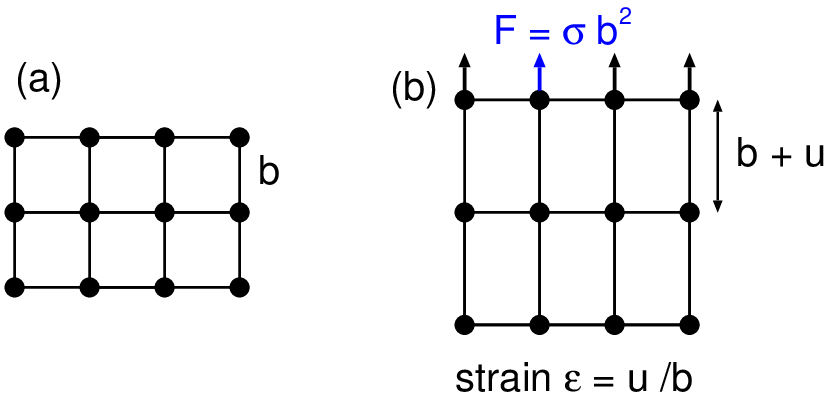}
\caption{\label{IndealStrength.eps}
When a rectangular block is elongated by a stress $\sigma$ the bonds between the
atom stretch by $u$ so the strain $\epsilon = u /b$.
The bond breaks when the displacement $u$ is of order $b$.
The strain and 
stress at the point of bond fracture are $\epsilon_{\rm c} \approx 1$ 
and $\sigma_{\rm c} = E \epsilon_{\rm c} \approx E$, respectively.
}
\end{figure}

\vskip 0.3cm
{\bf 3 Crack propagation in a long slab--stress method}

The stress field in the vicinity of crack tips 
depends on the distance $r$ from the crack tip as $\sim r^{-1/2}$. 
This result is very general and holds for static cracks, moving cracks, and cracks in
viscoelastic solids. In ideal brittle solids for low crack-tip speeds
the singular stress field $\sim r^{-1/2}$ holds down to atomic distances from the singularity. 
For the crack shown in 
Fig. \ref{CrackBlockThree.eps} we expect the stress a distance
$\sim h_0$ in front of the crack tip to be of order the applied stress $\sigma_0$ so that
$$\sigma (r) \approx \sigma_0 \left ({h_0\over 2 r}\right )^{1/2}$$
In order to break the 
bonds at the crack tip the stress for $r \approx a_0$, where $a_0$ is an atomic
distance, must equal
the stress $\sigma_{\rm c}$ to break bonds so that
$$\sigma_{\rm c} \approx \sigma_0 \left ({h_0\over 2 a_0}\right )^{1/2}$$
or
$$ \sigma_0 \approx \sigma_{\rm c} \left ({2 a_0\over h_0}\right )^{1/2}\eqno(3)$$
For cohesive cracks, assuming no plastic deformation, 
$\sigma_{\rm c} \approx E$ (see Fig. \ref{IndealStrength.eps}). This gives
$$\sigma_0 \approx E \left ({2a_0 \over h_0}\right )^{1/2}\eqno(4)$$
Next note that the work per unit surface area to break the 
bonds is $G \approx E a_0$.
Using this result in (4) gives
$$\sigma_0 \approx  \left ({ 2 E^2 a_0 \over h_0}\right )^{1/2} \approx
\left ({ 2 G E \over h_0}\right )^{1/2} $$
which is the same result as (2). Note that this result is obtained only if the stress
at the crack tip varies as $\sim r^{-1/2}$, which support the statement that 
the stress close to the crack tip has this singular form.

\vskip 0.3cm
{\bf 4 Discussion}

We have shown that the energy and stress 
methods used above give the same results when applied to cohesive cracks and
assuming no plastic deformations. However, when applied to adhesive cracks they result in a paradox or mystery
as we will now show. Consider an interfacial crack and assume that (2) and (3) both are valid.
This imply
$$\sigma_0 = \left ( {2G E \over h_0}\right )^{1/2} = \sigma_{\rm c} \left ({2a_0\over h_0}\right )^{1/2} . \eqno(5)$$
From (5) we get
$$a_0 = {GE \over \sigma_{\rm c}^2}\eqno(6)$$
If $F_{\rm c}$ is the force to break an adhesive bond and $n_0=1/b^2$ 
the concentration of bonds then $\sigma_{\rm c} = F_{\rm c} /b^2$.
Since $F_{\rm c} \approx G b^2 /a$, where $a$ is an atomic
length, we get $\sigma_{\rm c} \approx G/a$ and
from (6)
$$a_0 = {GE \over \sigma_{\rm c}^2}\approx {E a^2 \over G}\eqno(7)$$

Now the force to break an adhesive bond is in general 
independent of the elastic modulus $E$, and it is clear
from (6) or (7) that $a_0$ {\it in general 
cannot be an atomic distance} since the Young's modulus $E$
can take a wide range of values.
We will illustrate the problem with two practical cases. 

Consider first an adhesive crack between two flat diamond (111) surfaces 
where the dangling bonds are passivized 
by hydrogen atoms [self mated C(111)$(1\times 1)$H].
The work of adhesion for this system is $G\approx 0.2 \ {\rm J/m^2}$ (see Ref. \cite{carp}), 
the Young's modulus of diamond
$E\approx 1.2\times 10^{12} \ {\rm Pa}$ and using the atomic distance
$a\approx 0.3 \ {\rm nm}$ from (7) we get
$$a_0 \approx {E a^2 \over G} \approx 5  \ {\rm \mu m}$$
Hence the bond-breaking occur a long distance away from the point
where the stress-singularity occur in the continuum approximation. We attribute this to the large modulus of diamond and
small work of adhesion of the passivized diamond surface which will make the diamond surfaces nearly parallel, and separated by an
atomic distance, in a large region behind the crack tip.
This makes the point where a bond break somewhat undefined.
In any case, the fact that $a_0$ is large is not associated with any physical problem.

One could argue that since there is no direct way to determine $a_0$ the energy method of Sec. 2 is better than
the stress method of Sec. 3. However, both approaches depend on only one quantity which in most cases need to be determined
experimentally. In the energy approach this is the crack-energy $G$ (or rather $EG$) and in the stress approach the parameter
$\sigma_{\rm c} \surd a_0$. These parameters could both be determined experimentally 
using a simple set-up like the long-slab configuration studied above.
With $G$ or $\sigma_{\rm c} \surd a_0$ known, crack propagation in more complex situations could be studied as usual by 
calculating the relevant stress intensity factors. In some cases, as for the hydrogen-passivized diamond surfaces, it is possible to
calculate (or estimate) $G$ from first principle using quantum mechanical 
methods (see Ref. \cite{carp}), while it is less clear how to 
determine $\sigma_{\rm c} \surd a_0$ theoretically if not obtained indirectly from (6).

\begin{figure}
\centering
\includegraphics[width=0.95\columnwidth]{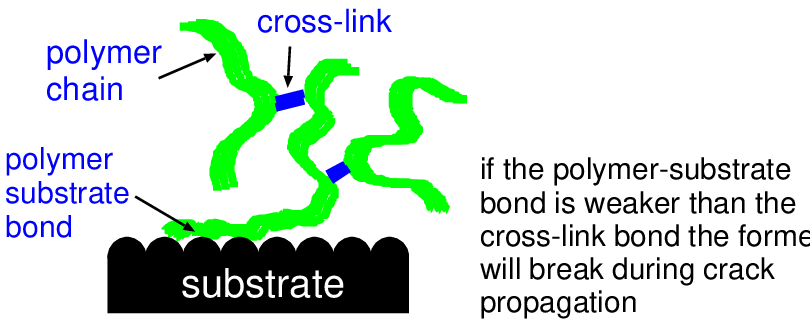}
\caption{\label{TwoBonds.eps}
Adhesive bonds are often of the weak Van der Waals type.
In these cases, for a covalent crosslinked polymer, the adhesion bonds will break before the 
chain or crosslinks break.
}
\end{figure}

Next, consider a rubber block adhering to a flat and rigid substrate.
As an example consider first a fully cross-linked Polydimethylsiloxane (PDMS, Sylgard 184) rubber
adhering to a glass surface. In this case
$E\approx 2\times 10^6 \ {\rm Pa}$ and $G\approx 0.1 \ {\rm J/m^2}$
(see, e.g., Ref. \cite{one,two}) and (7) gives
$$a_0 \approx {E a^2 \over G} \approx 10^{-13} \ {\rm m} . $$
Hence $a_0$ is much smaller than an atomic length which is impossible. To discuss this paradox first note that
using $a_0 = 0.3 \ {\rm nm}$ in (6) the stress at the crack tip
$$\sigma_{\rm c} = \left (G E \over a_0 \right )^{1/2} \approx 26 \ {\rm MPa}$$
In linear elasticity with $E=  2\times 10^6 \ {\rm Pa}$ this correspond to the strain $\epsilon \approx \sigma_{\rm c}/E \approx
13$ and it is clear that linear elasticity theory will fail close to the crack tip even for the relative weak adhesion considered
here. The stress-strain curve for large strain for PDMS is (highly) nonlinear with the stress increasing much faster than linear with increasing strain
as the polymer chains approach their fully stretched state. For PDMS Sylgard 184 the yield stress in tension for the perfect (defect free)
block may be larger than the critical stress $\sigma_{\rm c} \approx 26 \ {\rm MPa}$ found above, in which case the rubber at the crack tip 
would respond as a nonlinear elastic material but without breaking crosslinks and without 
formation of defects such as cavities. This conclusion is also supported by the result illustrated in
Fig. \ref{TwoBonds.eps}: as long as the substrate bond (which typically is of the Van der Waals nature, and hence
very weak) is weaker than the cross-link bonds one expect the polymer-substrate bond to break before the cross-link bonds break.
However, for weakly crosslinked PDMS the elastic modulus is much smaller (e.g. only $19 \ {\rm kPa}$ for Sylgard PDMS 
produced by mixing cross-linker/base fluids in the ratio 1:50) 
and in these cases the strain $\sigma_{\rm c}/E$ calculated from (6) with $a_0$ as an atomic
distance would be huge and the rubber would break-up (e.g. cavitation and stringing) close to the crack tip.
(In addition, weakly crosslinked rubber have free chains which could get 
pulled out during opening crack propagation\cite{chain1,chain2,chain3}. This has been observed experimentally for the contact between a silica glass sphere and
PDMS rubber. Here the effective work of adhesion decreases with the number of contacts due to pullout and 
transfer of uncrosslinked chains to the glass surface\cite{one,pullout}.) 

The results presented above shows that if $a_0$ as calculated from (7) is smaller than an atomic length the linear elastic theory cannot be used
all the way down to atomic distances from the crack tip but depending on the system one need to take into account 

(a) non-linear elasticity effects close to the crack tip, which may effectively increase the elastic modulus $E$, and hence also $a_0$, and

(b) plastic deformation, or the formation of defects, such as cavities or (for polymers) stringing, in a region close to the crack tip.   

In both cases the effective $a_0$ would be larger than predicted by (7).

We note that for elastic solids the cut-off distance $a_0$ and $\sigma_{\rm c}$ always enter in the combination
$\sigma_{\rm c} \surd a_0$ and it is usually not possible to measure $\sigma_{\rm c}$ and $a_0$ separately. The analysis above shows that 
if one choose $\sigma_{\rm c}^2 a_0 = G_0$ the stress-theory will give the same result as the energy-theory. However,
for viscoelastic solids (here denoted as rubber) the situation is different. For very low sliding speeds (or high temperature) the rubber 
is in the rubbery region and behaves as an elastic solid and the crack propagation will only depend on the (low velocity) crack-energy 
$G_0 = G(v\rightarrow 0)$ which we again can write as $\sigma_{\rm c} \surd a_0$. However, as we increase the sliding speed there will be viscoelastic energy dissipation
in a region around the crack tip. A crack tip which moves with the speed $v$ will generate time dependent deformations of the rubber
with a band of frequencies, $\omega \sim v/r$ ($a < r < \infty$), where $r$ is the distance from the crack tip.
The highest frequency component is given by $v/a$, where $a=a(v)$ is the crack-tip cut-off length with $a\rightarrow a_0$ as $v\rightarrow 0$.
Hence the crack-energy $G(v)$ will now depend on {\it both} $G_0$ and $a_0$. The fracture energy $G_0$ can be directly measured
experimentally, but the cut-off distance $a_0$ is not so easy to measure directly but could be deduced by comparing the measured $G(v)$ curve to 
the theory prediction (note: the $G(v)$-curve shift along the velocity axis linearly with $a_0$). But as shown above, if linear viscoelasticity
is used this could result in cut-off length $a_0$ which would be unphysical small. This would signal that non-linear viscoelastic effects,
or other phenomena such as cavitation, occur close to the crack tip. 

For adhesive cracks, the stress field far from the crack tip, where most of the viscoelastic energy dissipation occur, 
may be small enough for linear viscoelasticiy to be valid. But when calculating (or estimating) the
viscoelastic contribution to the crack-energy for rubber-like materials one must use for $a_0$ a length at least of order some monomers 
(say $\sim 1 \ {\rm nm}$). 

\begin{figure}[tbp]
\includegraphics[width=0.35\textwidth,angle=0]{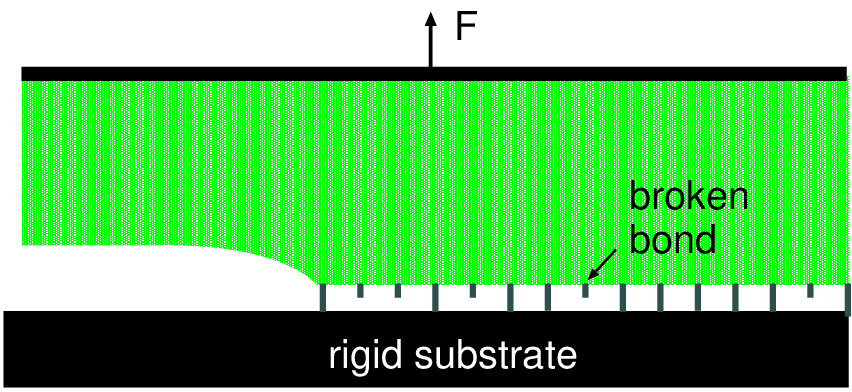}
\caption{
For adhesive crack propagation a fraction of the bonds will be broken by thermal fluctuations even far in front of the
crack tip. The concentration of broken bonds increases as the crack tip is approached from the bonded side.
}
\label{CrackPull.eps}
\end{figure}

\vskip 0.3cm
{\bf 5 Thermal effects on crack propagation}

Theories for cracks in solids often assume a crack tip process zone
which can be very complex involving plastic flow, cavitation, stringing and bond-breaking far from the crack tip. 
However, here we consider adhesive crack propagation and we 
assume that the stress close to the crack tip is so small that 
only the (weak) interfacial bonds break. For weak bonds
thermal activation can be very important and can result in the interfacial 
bonds breaking (and reforming) already far in front of the crack tip. 

If the elastic energy $U_{\rm el} > U_{\rm ad}$ opening crack 
propagation occur while if $U_{\rm el} < U_{\rm ad}$ closing crack propagation
occur. For zero temperature if $U_{\rm el} = U_{\rm ad}$ a crack will not move (stationary crack). However, for $T>0$
cracks can move by thermal excitation even if $U_{\rm el} = U_{\rm ad}$.
In this case, the crack length could both increase or decrease,
correspond to opening and closing crack propagation, and in general the crack tip would perform a
random, Brownian-motion type of movement\cite{random}. 
However, in many practical situations when bonds break some irreversible
process occur, such as passivization of the dangling bonds by reaction with air or contamination 
molecules, or surface reconstruction, and in these cases only opening crack propagation may occur.
This could result in a slow increase in the crack length even when $U_{\rm el} < U_{\rm ad}$ 
as observed in stress corrosion cracking. We will study this problem in another publication.

A crack is a line defects and thermal excitation's will result in 
different regions of a crack-line moving at different times
resulting in a curved crack-line.
Thus the energy barrier for movement of the crack-line as a result of 
thermal excitation will not
involve the crack-line moving forwards uniformly, 
which for an infinite long crack-line 
would involve an infinite large energetic barrier (as infinite many bonds would need to be broken at the same time),
but small nano-sized regions will move nearly randomly in time. Thus the energy barrier for crack propagation discussed below
correspond to the barrier for the motion of a nano-sized crack-line segment.

Consider a crack propagating with a constant speed $v$ in an infinite 
viscoelastic solid (see Fig. \ref{CrackPull.eps}). The adhesive (or cohesive) bonds on 
the surface $z=0$ in front of a crack tip can break before the stress
from the crack reach the critical value $\sigma_{\rm c}$ 
where the stress alone would break the bonds.
This is due to thermal fluctuations which can supply part of the energy needed
to break a bond. Thermal effects are particularly important for weak bonds, e.g.,
Van der Waals bonds.

We will denote a bond in front of the crack tip to be broken by the propagating crack
as an interfacial bond. Let $P(t)$ be the probability that an interfacial bond is 
not broken at time $t$. We assume the stress singularity in the continuum mechanics
formalism occur a distance $a$ behind the opening crack tip so the stress at the crack tip
is finite. Let $x=vt$ (with $t \le 0$) 
be the position of the crack tip defined so that the 
stress on the surface $z=0$ for $x>vt$ is given to leading order by
$$\sigma_{zz} = {K_{\rm I}\over 2 \pi (x+a-vt)^{1/2}}\eqno(8)$$
where $K_{\rm I}$ is the stress intensity factor for opening crack propagation. 
We denote $\sigma_{zz} = \sigma (t)$ for simplicity. In (8) we have
assumed that the crack-line is strait. 

\begin{figure}[tbp]
\includegraphics[width=0.47\textwidth,angle=0]{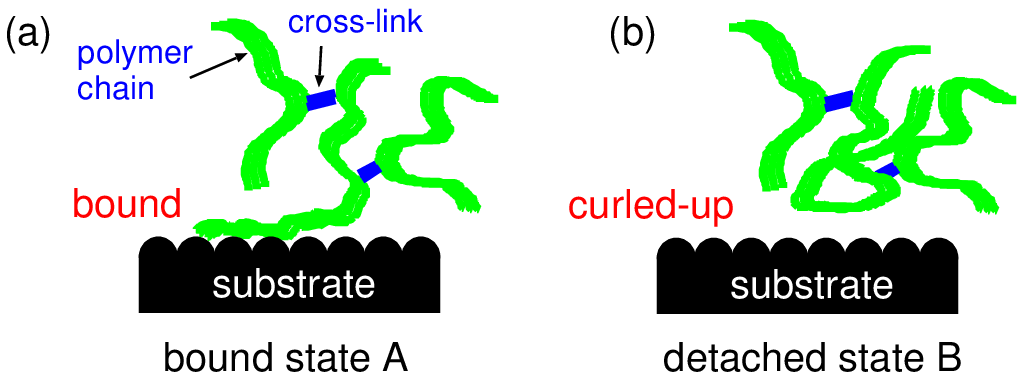}
\caption{
A polymer chain in a bound state (a) and in the detached state (b). In the detached state the chain
is ``curled-up'' and an activation barrier is $V_{\rm form}$ must be overcome to form a new bonded state.
The barrier $V_{\rm break}$ to break the bond is larger than $V_{\rm form}$ and the chain spend most of the time in the bound state.
}
\label{TwoBonds1.eps}
\end{figure}

Consider now an interfacial bond Q and 
choose the origin of the $x$-axis so that the 
interfacial bond Q (broken or not broken)
occupy the position $x=0$ at $t=0$. 
Let $F_{\rm c}$ denote the force to break the bond 
at zero temperature. For $T>0$ the bond will break before the external force reaches $F_{\rm c}$. 
This is due to thermal fluctuations, which can supply the 
energy needed to break a bond before the external force reach the critical value
$F_{\rm c}$. 

\begin{figure}[tbp]
\includegraphics[width=0.47\textwidth,angle=0]{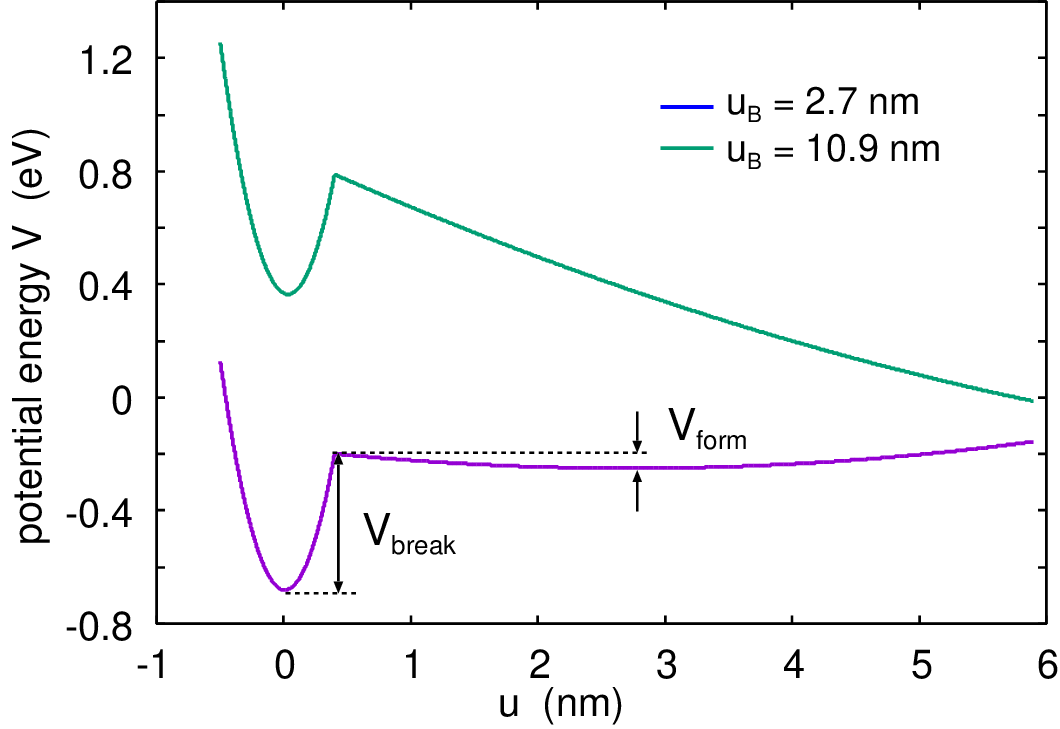}
\caption{
The potential energy $V(u)$ [given by (11)-(13)] when the surface separation $u_{\rm B} = 2.7$ and $10.9 \ {\rm nm}$.
For $u_{\rm B} =10.9 \ {\rm nm}$ the potential well of the detached state occur for larger $u$ than shown in the figure.
For the parameters $U_{\rm A0}=0.5 \ {\rm eV}$, $U_{\rm B}=0.25 \ {\rm eV}$, $u_{\rm c} = 0.4 \ {\rm nm}$,
$k_{\rm B} = 0.003 \ {\rm N/m}$ and $k_{\rm A} = 2 U_{\rm A}/u_{\rm c}^2 = 1.0 \ {\rm N/m}$.
}
\label{1u.2V.1.and.4.eps}
\end{figure}

We consider the case where a broken bond can reform again. 
This will almost always be the case for weak
bonds such as the Van der Waals bonds often involved in rubber adhering to a rigid substrate. In this case a polymer chain at the interface can
be in a state A where it binds to the substrate or in a detached (curled-up) state B as illustrated in Fig. \ref{TwoBonds1.eps}. 
In the weakly 
stretched state 
far in front of the crack tip an interfacial polymer chain is most of the time 
in the bound state A 
but sometimes in a detached state B. We assume the potential
energy has local minima for the bound state and for the detached state (see Fig. \ref{1u.2V.1.and.4.eps}). The probability that
the chain is in the bound state is denoted by $P_{\rm A}$ and in the detached state with $P_{\rm B}$. We assume that $P_{\rm A}(t)$ and $P_{\rm B}(t)$ obey the
rate equation
$${d P_{\rm A} \over dt} = - \kappa_{\rm A} P_{\rm A} + \kappa_{\rm B} P_{\rm B}\eqno(9)$$
where $P_{\rm A}+P_{\rm B}=1$ is time independent. 
We assume that
$$\kappa_{\rm A} = \nu_{\rm A} e^{-\beta V_{\rm break}}, \ \ \ \ \ \kappa_{\rm B} = \nu_{\rm B} e^{-\beta V_{\rm form}}\eqno(10)$$
where the barrier heights $V_{\rm break}$ and $V_{\rm form}$ 
are defined in Fig. \ref{1u.2V.1.and.4.eps} and depends on time due to the external
force $F(t)$ stretching the chain. We assume (to be discussed below) that 
$\nu_{\rm A}=\nu_{\rm B} = k_{\rm B} T /h$, where $h$ is the Planck constant.
This is the Eyring expressions for pre-exponential factors $\nu_{\rm A}$ and $\nu_{\rm B}$.

Consider an interfacial bond Q as the crack tip approach it.
The local stress at bond Q is given by (8).
We assume that the potential energy of a bond is given by
$$V(u)=V_{\rm A}(u)+V_{\rm B}(u)\eqno(11)$$
where
$$V_{\rm A}(u)= -U_{\rm A} +{1\over 2} k_{\rm A} u^2\eqno(12)$$
for $u<u_{\rm c}$ and $V(u)=0$ for $u>u_{\rm c}$ which correspond to a broken 
molecule-substrate bond, and
$$V_{\rm B}=- U_{\rm B} +{1\over 2} k_{\rm B} (u-u_{\rm B})^2\eqno(13)$$
which is the energy of a stretched polymer chain. The polymer chain does not break so this is
the energy of the chain for all chain (extension) length $u$. 

The adiabatic (infinitely slowly) work to break a bond is the difference in the potential
energy after and before breaking the bond. The potential energy at the equilibrium state (with
$u=u_{\rm B}=0$) is $-U_{\rm A}-U_{\rm B}$ and for the state with broken bond $-U_{\rm B}$. Thus, 
the (adiabatic) work of adhesion $w_0 = U_{\rm B}/b^2$ [which equals $G(v\rightarrow 0)$],
where $b^2$ is the area occupied by the bonding segment of the polymer chain.

The chain-substrate bond break when
$$u=u_{\rm c}= \left ({2 U_{\rm A}\over k_{\rm A}}\right )^{1/2}\eqno(14)$$ 
where the potential energy $V=V_1$ with
$$V_1 = V_{\rm B}(u_{\rm c}) =  - U_{\rm B} +{1\over 2} k_{\rm B} (u_{\rm c}-u_{\rm B})^2\eqno(15)$$
Before the bond is broken the potential energy $V(u)$ is minimal when $V'(u)=0$ or
$$k_{\rm A} u+k_{\rm B} (u-u_{\rm B})=0$$
or 
$$k_{\rm A} u= K u_{\rm B}\eqno(16)$$
where
$$ K={k_{\rm A} k_{\rm B} \over k_{\rm A}+k_{\rm B}}\eqno(17)$$
Using (12), (13) and (16) in (11) gives the potential energy $V=V_0$ with
$$V_0 = -(U_{\rm A}+U_{\rm B})+ {1\over 2} K u_{\rm B}^2\eqno(18)$$ 

Thus the effective barrier for bond breaking is
$$V_{\rm break} = V_1-V_0 = U_{\rm A}+{1\over 2} k_{\rm B} (u_{\rm c}-u_{\rm B})^2-{1\over 2} K u_{\rm B}^2$$ 
$$=U_{\rm A}+{1\over 2} k_{\rm B} \left (u_{\rm c}^2-2u_{\rm c} u_{\rm B} + {k_{\rm B}\over k_{\rm A}+k_{\rm B}} u_{\rm B}^2 \right )\eqno(19)$$
The energy barrier to form a bond when the surface separation equals $u_{\rm B}$ is 
$$V_{\rm form} = V_1-[-U_{\rm B}] = {1\over 2} k_{\rm B} (u_{\rm c}-u_{\rm B})^2\eqno(20)$$
Note that when $k_{\rm B} = 0$ then $V_{\rm break} = U_{\rm A}$ and $V_{\rm form}=0$ as expected.

If an external force $F$ stretch the bond then
$$k_{\rm A} u = k_{\rm B} (u_{\rm B}-u) = F$$
so the displacement $u=u_F = F/k_{\rm A}$ and 
$$u_{\rm B} = {F\over K} . \eqno(21)$$

We denote the force to break the bond at zero temperature by $F_{\rm c} = k_{\rm A} u_{\rm c}$
The displacement $u_{\rm B}=u_{\rm Bc}$ when the bond break (at zero temperature) is $u_{\rm Bc} = F_{\rm c}/K$
which we can also write as
$$u_{\rm Bc} = {k_{\rm A}+k_{\rm B}\over k_{\rm B}} u_{\rm c}$$
Substituting $u_{\rm B} = F/K$ in (19) and (20) gives the barriers 
$V_{\rm break}(F)$ and $V_{\rm form}(F)$ as a function of the applied force $F$.

Using $P_{\rm B}=1-P_{\rm A}$ from (9) we get
$${d P_{\rm A} \over dt} = - \kappa P_{\rm A} + \kappa_1\eqno(22)$$
where $\kappa=\kappa_{\rm A}+\kappa_{\rm B}$ and $\kappa_1 = \kappa_{\rm BA}$. In what follows we denote $P_{\rm A}= P$ for simplicity.
Far in front of the crack tip $dP/dt=0$ so that $P = P_{0}$ where
$$P_{0} = {\kappa_1 \over \kappa} = {1 \over 1+e^{-\beta \Delta E_0}}$$
where $\Delta E_0 = V_{\rm break}-V_{\rm form}$ is the energy difference between the two potential wells.

From (21) we get
$$P(t) = P(-\infty) {\rm exp}\left ( {-\int_{-\infty}^t dt' \kappa(t')}\right )$$
$$ + \int_{-\infty}^t dt' \ \kappa_1(t') {\rm exp} \left ({-\int_{t'}^t dt'' \kappa(t'')}\right )\eqno(23)$$

\vskip 0.3cm
{\bf 6 Dependency of the crack-energy on the crack-tip speed and the temperature}

Consider now a molecular bond at $x=0$ as the crack tip approach it from $t=-\infty$. 
A bond experience the force
$$F(t) = {K_{\rm I} b^2 \over 2 \pi (a-vt)^{1/2} },$$
as the crack tip approach it. We choose $a$ so that $F(0)=F_{\rm c}$ so that
$$F(t) = F_c \left ({ a \over a-vt }\right )^{1/2}\eqno(24)$$
where
$$F_c = {K_{\rm I} b^2 \over 2 \pi \surd a}\eqno(25)$$
The crack-energy $G$ is related to $K_{\rm I}$ using\cite{standard1,standard2}
$$G={K_{\rm I}^2 \over E}\eqno(26)$$
which gives
$$G=\left ({F_c 2 \pi \over b^2 } \right )^2 {a\over E} \eqno(27)$$
The crack-energy as $v\rightarrow 0$ is given by
$$G_0 \approx {U_{\rm A} \over b^2} = {1\over 2 b^2} k_{\rm A} u_c^2 = {1\over 2 b^2 k_{\rm A} } F_c^2\eqno(28)$$
so we can write
$$G= {a\over a_0} G_0\eqno(29)$$
where
$$a_0 = {b^2 E \over 8 \pi^2 k_{\rm A}}\eqno(30)$$

The work per unit area to propagate the crack is
$$G= {1\over b^2} \int_{-\infty}^0 dt \ \dot u_{\rm B} (t) F(t) P(t)$$
If we change integration variable to $x=a-vt$ this gives
$$G= {1\over b^2} \int_{a}^\infty dx \ \left [-{du_{\rm B} \over dx}\right ] \  F(x) P(x)\eqno(31)$$
and from (23) with $dt = - dx/v$
$$P(x) = P(\infty) {\rm exp}\left (-{1 \over v} \int_{x}^{\infty} dx' \ \kappa (x') \right )$$
$$ + {1\over v} \int_{x}^\infty dx' \ \kappa_1(x') 
{\rm exp} \left (-{1 \over v} \int_{x}^{x'} dx''  \ \kappa(x'') \right )\eqno(32)$$ 
For any finite $x$ the first integral in (32) will vanish because $\kappa (x)$ approach a constant
finite value for large $x$. Hence we get
$$P(x) = {1\over v} \int_{x}^\infty dx' \ \kappa_1(x') 
{\rm exp} \left (-{1 \over v} \int_{x}^{x'} dx''  \ \kappa(x'') \right )\eqno(33)$$ 
From (24)
$$F= F_c \left ({ a\over x}\right )^{1/2}\eqno(34)$$
and using that $F =K u_{\rm B} $ this gives
$$-{du_{\rm B} \over dx} = -{1\over K} {dF \over dx} = {F_c \over 2 K} {a^{1/2} \over x^{3/2}}$$
Thus (31) becomes
$$G= {F_c^2\over 2 K b^2} \int_{a}^\infty dx \ {a \over x^2} P(x)\eqno(35)$$

At zero temperature $P(x) = 1$ for $x>0$ so that the crack propagation energy $G=G_1$ for $T=0$ is
$$G_1 = {F_c^2 \over 2 K b^2} \int_{a}^\infty dx \ {a\over x^2} = 
{F_c^2 \over 2Kb^2} \eqno(36)$$
Using that $K u_{\rm Bc} = F_c$ this can also be written as $F_c u_{\rm Bc} /2b^2$.
Using (35) and (36) we get
$${G\over G_1}=  \int_{a}^\infty dx \ {a \over x^2} P(x)\eqno(37)$$

Let us consider the limits $v\rightarrow 0$ and $v\rightarrow \infty$.
As $v \rightarrow \infty$ the integrals in (33) will be dominated by the
large $x'$ and $x''$ regions and we can replace $\kappa_1 (x')$ and
$\kappa (x'')$ with their values for large $x$ which we denote by
$\kappa_{\rm 1eq}$ and $\kappa_{\rm eq}$ since these are the
thermal equilibrium rate constants. We get
$$P(x) \approx 
{1\over v} \kappa_{\rm 1eq} \int_{x}^{\infty} dx' \ 
{\rm exp} \left (-{1 \over v} \kappa_{\rm eq} (x'-x)\right )
= {\kappa_{\rm 1eq} \over \kappa_{\rm eq}}\eqno(38)$$
which is equal to the probability
$P_{\rm eq} = \kappa_{\rm 1eq}/ \kappa_{\rm eq}$
that an interfacial molecule is in the
binding state $A$ far in front of the crack tip.
Using (37) we get $G\rightarrow G_\infty$ as $v\rightarrow \infty$ where
$${G_\infty \over G_1} =  \int_{a}^\infty dx \ {a\over x^2} 
{\kappa_{\rm 1eq}\over \kappa_{\rm eq}} = {\kappa_{\rm 1 eq}\over \kappa_{\rm eq}}=P_0$$
In the numerical study presented below $P_0\approx 1$.

Next consider the limit $v\rightarrow 0$. In that case only a small
$[x'-x]$-interval will contribute to the inner integral in (33) and we can
replace $\kappa (x'')$ with its value for $x''=x$. This gives
$$P(x) \approx 
{1\over v} \int_{x}^{\infty} dx' \ \kappa_1 (x')
{\rm exp} \left (-{1 \over v} \kappa (x) (x'-x)\right )$$
For $v\rightarrow 0$ only $x'$ very close to $x$ will contribute to this integral and we
can replace $\kappa_1 (x')$ with $\kappa_1 (x)$ to get
$$P(x) \approx {\kappa_1 (x) \over \kappa (x)}$$
Using this in (37) we get $G\rightarrow G_0$ as $v\rightarrow 0$ where
$${G_0\over G_1} =  \int_{a}^\infty dx \ {a\over x^2} {\kappa_1(x)\over \kappa(x)}$$
Writing $\xi = a/x$ this gives
$${G_0\over G_1} =  \int_{0}^1 d\xi \ {\kappa_1(\xi)\over \kappa(\xi)}$$

Changing integration variable in (37) and (32) from $x$ to $\xi = a/x$ gives
$${G\over G_1} =  \int_{0}^1 d\xi \ P(\xi)\eqno(39)$$
and 
$$P(\xi ) = 
{a\over v} \int_{0}^{\xi} d\xi' \  {\kappa_1(\xi') \over [\xi']^2} 
{\rm exp} \left (-{a \over v} \int_{\xi'}^{\xi} d\xi''  \ {\kappa(\xi'')\over [\xi'']^2} \right )\eqno(40)$$ 

Since $\kappa$ and $\kappa_1$ are functions of $u_{\rm B}$ we need to relate $u_{\rm B}$ to $\xi =a/x$.
Using (34) we get
$$u_{\rm B} = {F\over K} = {F_c\over K}\left ({a\over x}\right )^{1/2} = {F_c\over K}\xi^{1/2}$$

For numerical calculations the $\xi''$-integral in (40) is problematic as $v\rightarrow 0$
where only a very small $\xi-\xi'$ interval will effectively contribute. This problem is solved by using
as integration variable instead of $\xi''$ the variable $\eta$ defined by $\xi''= \xi e^{-\eta}$ giving
$d\xi'' = - \xi'' d\eta$ and
$$\int_{\xi'}^{\xi} d\xi''  \ {\kappa(\xi'')\over [\xi'']^2}  = \int_0^{\eta_0} d\eta \ {\kappa(\xi'')\over \xi''}$$ 
where $\eta_0 = {\rm log} (\xi/\xi')$.

\begin{figure}[tbp]
\includegraphics[width=0.47\textwidth,angle=0]{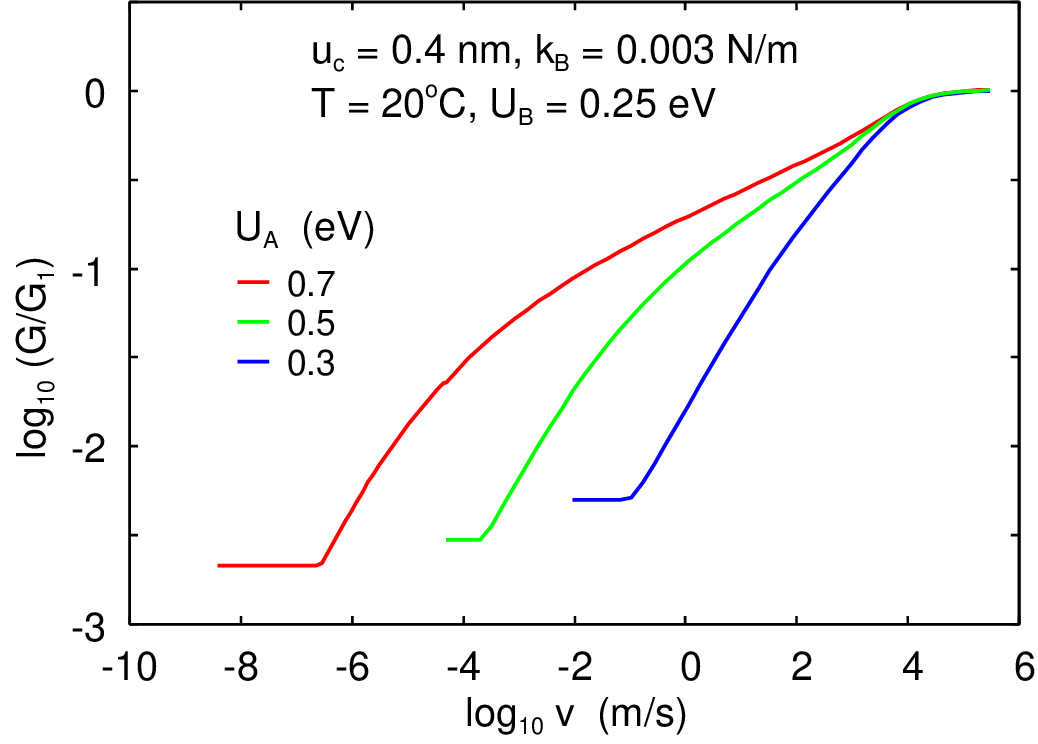}
\caption{
The crack-energy $G$ in units of its zero temperature value $G_1=G(T=0)$
as a function of the crack-tip speed (Log-Log scale) for the bond energies
$U_{\rm A} = 0.3$, $0.5$ and $0.7 \ {\rm eV}$. For $T=20^\circ {\rm C}$, the 
spring constant $k_{\rm B}=0.003 \ {\rm N/m}$, the separation to break the bond $u_{\rm c} = 0.4 \ {\rm nm}$,
and the bond energy $U_{\rm B}=0.25 \ {\rm eV}$. 
For the three cases with $U_{\rm A}=0.3$, $0.5$ and $0.7 \ {\rm eV}$ we have
$G_1 b^2 = 60.4$, $167.4$ and $327.8 \ {\rm eV}$.
}
\label{1logv.2logG.over.GT=0.eps}
\end{figure}

\begin{figure}[tbp]
\includegraphics[width=0.47\textwidth,angle=0]{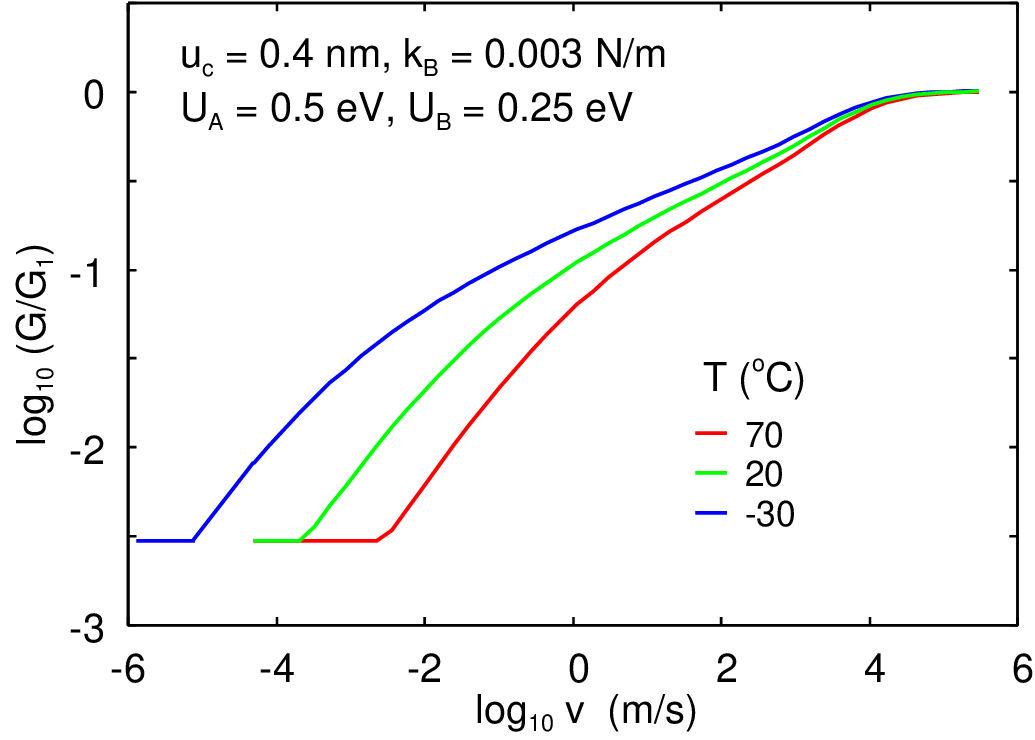}
\caption{
The crack-energy $G$ in units of its zero temperature value $G_1=G(T=0)$
as a function of the crack-tip speed (Log-Log scale) for the temperatures
$T=-30$, $20$ and $70^\circ {\rm C}$
The spring constants
$k_{\rm B} = 0.003$. For the 
A-well bond energy $U_{\rm A}=0.5 \ {\rm eV}$, the separation to break the bond $u_{\rm c} = 0.4 \ {\rm nm}$,
and the B-well bond energy $U_{\rm B}=0.25 \ {\rm eV}$. 
}
\label{1logv.2logG.varyT.eps}
\end{figure}

\begin{figure}[tbp]
\includegraphics[width=0.47\textwidth,angle=0]{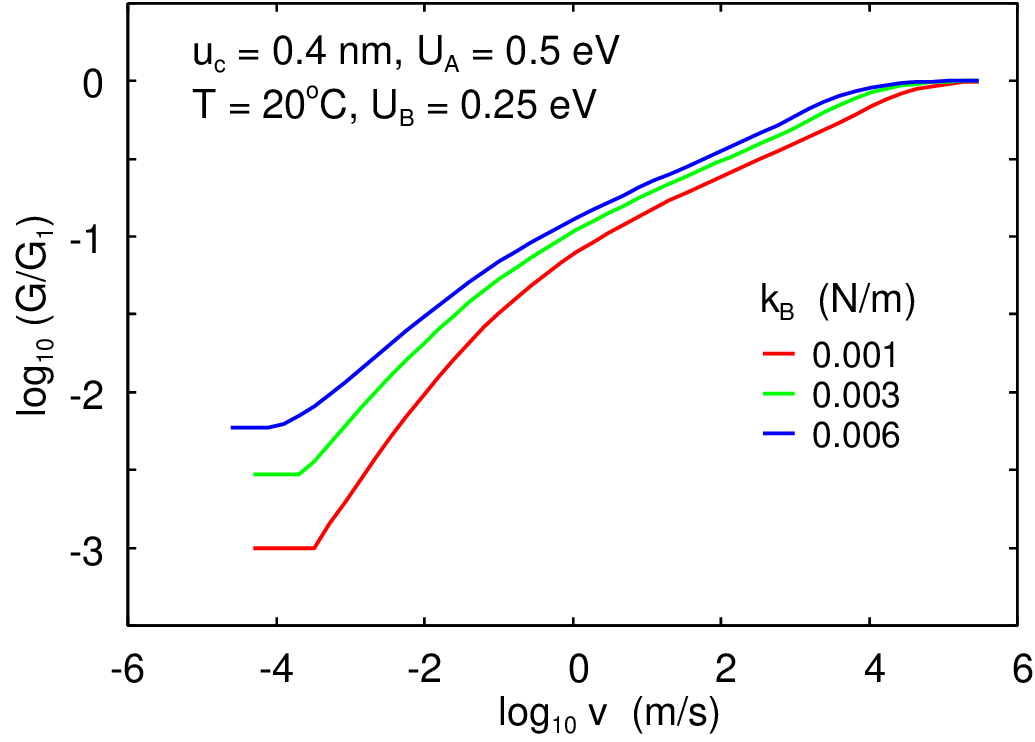}
\caption{
The crack-energy $G$ in units of its zero temperature value $G_1=G(T=0)$
as a function of the crack-tip speed (Log-Log scale) for the spring constants
$k_{\rm B} = 0.001$, $0.0.003$ and $0.006 \ {\rm N/m}$. For $T=20^\circ {\rm C}$, the 
A-well bond energy $U_{\rm A}=0.5 \ {\rm eV}$, the separation to break the bond $u_{\rm c} = 0.4 \ {\rm nm}$,
and the B-well bond energy $U_{\rm B}=0.25 \ {\rm eV}$. 
For the three cases with $k_{\rm B} = 0.001$, $0.0.003$ and $0.006 \ {\rm N/m}$ we have
$G_1 b^2 = 501.2$, $167.4$ and $84.0 \ {\rm eV}$.
}
\label{1logv.2logG.over.G1.vary.kB.eps}
\end{figure}

\vskip 0.3cm
{\bf  Numerical results and discussion}

The theory for $G(v,T)$ presented above depends on a set of parameters,
the most important being $U_{\rm A}$, $k_{\rm B}$ and the temperature $T$. Here we will present
numerical results illustrating this. The reference curve (green curves in Fig. \ref{1logv.2logG.over.GT=0.eps}-\ref{1logv.2logG.varyT.eps})  
is obtained using $T_0=20^\circ {\rm C}$, $U_{\rm A} = 0.5 \ {\rm eV}$, $U_{\rm B}=0.25 \ {\rm eV}$,
$k_{\rm B}=0.003 \ {\rm N/m}$ and $u_c = 0.4 \ {\rm nm}$. 

Fig. \ref{1logv.2logG.over.GT=0.eps}
shows the crack-energy $G$ in units of its zero temperature value $G_1=G(T=0)$
as a function of the crack-tip speed (Log-Log scale) for the bond energies
$U_{\rm A} = 0.3$, $0.5$ and $0.7 \ {\rm eV}$, and the other parameters as for the reference case. 
For $U_{\rm A}=0.3$, $0.5$ and $0.7 \ {\rm eV}$ we have
$G_1 b^2 = 60.4$, $167.4$ and $327.8 \ {\rm eV}$, respectively.

Fig. \ref{1logv.2logG.varyT.eps}
shows similar results as in Fig. \ref{1logv.2logG.over.GT=0.eps} for the temperatures
$T=-30$, $20$ and $70^\circ {\rm C}$,
and Fig. \ref{1logv.2logG.over.G1.vary.kB.eps} for the spring constants,
$k_{\rm B} = 0.001$, $0.0.003$ and $0.006 \ {\rm N/m}$,
and the other parameters as for the reference case. 
For $k_{\rm B} = 0.001$, $0.0.003$ and $0.006 \ {\rm N/m}$ we have
$G_1 b^2 = 501.2$, $167.4$ and $84.0 \ {\rm eV}$.

\begin{figure}[tbp]
\includegraphics[width=0.47\textwidth,angle=0]{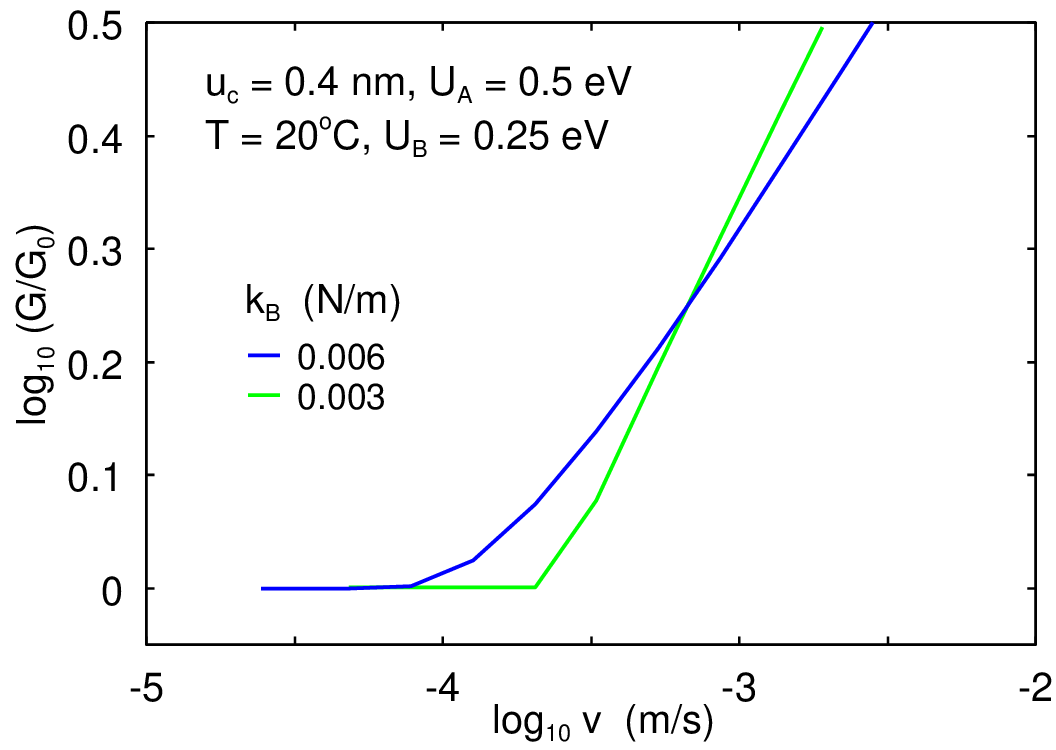}
\caption{
The crack-energy $G$ in units of its low-velocity value $G_0=G(v=0)$
as a function of the crack-tip speed (Log-Log scale) for the spring constants
$k_{\rm B} = 0.001$ and $0.006 \ {\rm N/m}$. For $T=20^\circ {\rm C}$, the 
A-well bond energy $U_{\rm A}=0.5 \ {\rm eV}$, 
the separation to break the bond $u_{\rm c} = 0.4 \ {\rm nm}$,
and the B-well bond energy $U_{\rm B}=0.25 \ {\rm eV}$. 
For the two cases with $k_{\rm B} = 0.001$ and $0.006 \ {\rm N/m}$ we have
$G_1 b^2 = 501.2$ and $84.0 \ {\rm eV}$.
}
\label{1logv.2logG.overG0.eps}
\end{figure}

The numerical results in Fig. \ref{1logv.2logG.over.GT=0.eps}-\ref{1logv.2logG.varyT.eps} 
show a low velocity range where $G_0(v)$ is velocity independent followed by a very
abrupt increase in $G_0(v)$ at a characteristic speed $v_{\rm c}$ which depends on the binding energy $U_{\rm A}$, the spring
constant $k_{\rm B}$ and the temperature $T$. In the low velocity ``adiabatic'' region $v<v_{\rm c}$ the separation of the surfaces at the
crack tip is so slow that at any moment in time the molecules are in dynamical (kinetic) equilibrium between the bonding state A and 
the detached state B. That is, during the separation an interfacial molecule oscillates  (randomly in time) 
between the states A and B in such a way that the ratio of the
probability to find the system in state B and A is given by the Boltzmann factor ${\rm exp}(-\beta \Delta E)$, 
where $\Delta E = V_{\rm break}-V_{\rm form}$ is the (separation dependent) energy difference between the two states. 
This is similar to the behavior of fluids at low shear rate. 

Fig. \ref{1logv.2logG.over.GT=0.eps} shows that the critical velocity $v_{\rm c}$ decreases as the binding energy $U_{\rm A}$ increases. 
For binding energies as large as expected for cohesive cracks ($\sim 3 \ {\rm eV}$) $v_{\rm c}$ will be so low that no
adiabatic velocity region can be observed on any time scale involving human activities (see Sec. 8). Nevertheless, $G_0(v)$ will depend on the velocity
and only for relative high velocities will $G_0(v)$ be velocity independent. The reason is that independent of the magnitude of
$U_{\rm A}$ when the external force has pulled the bond close to the point where the bond would be broken by the applied force alone, 
the remaining barrier for bond breaking will be overcome by a thermal fluctuation resulting in a velocity-dependent $G_0(v)$.
This is also clear from Fig. \ref{1logv.2logG.over.GT=0.eps} where for all the studied cases $G_0(v)$ reach the high-velocity
plateau for $\sim 10^4 \ {\rm m/s}$.

Fig. \ref{1logv.2logG.over.G1.vary.kB.eps} shows that variations in the spring constant $k_{\rm B}$ result in relative 
small variations in the critical velocity $v_{\rm c}$. However, the magnitude of the ratio between the high-velocity
$G_0(v)$ and of $G_0 (v)$ 
in the adiabatic velocity region increases strongly with decreasing $k_{\rm B}$. This is expected from
the Lake and Thomas description of the dependency of the crack-energy on the chain length (between cross links); decreasing
$k_{\rm B}$ correspond to increasing chain length $L$ ($k_{\rm B} \sim 1/L$).

Fig. \ref{1logv.2logG.varyT.eps} shows that as the temperature decreases $v_{\rm c}$ decreases. This is a trivial result
because thermal fluctuations decreases with decreasing temperatures and for $T\rightarrow 0$ we have 
$G\rightarrow G_1$ and $v_{\rm c} \rightarrow 0$. 

The abrupt increase in $G_0(v)$ at $v=v_{\rm c}$ observed in Fig.
\ref{1logv.2logG.over.GT=0.eps}, \ref{1logv.2logG.over.G1.vary.kB.eps}
and \ref{1logv.2logG.varyT.eps} is in good agreement with 
what we have found in adhesion experiments for PDMS in contact with glass surfaces,
see Fig. \ref{Combined.ps}. Using a reasonable binding energy $U_{\rm A} = 0.5 \ {\rm eV}$ (see introduction) and a reasonable spring constant of order
$0.003-0.006$ we get the onset of non-adiabatic effects at a critical velocity
$v_{\rm c}\approx 0.1 \ {\rm mm/s}$, in good agreement with experiments (see Fig. \ref{Combined.ps}).
Furthermore, in the velocity region studies $G_0(v)$ increases nearly linearly with $v$ on the Log-Log scale, both in the experiment and
in the theory. This indicate  $G_0 \sim v^\alpha$ with $\alpha \approx 0.22$ in the experiment (Fig. \ref{Combined.ps} 
and Ref. \cite{one}). The theory predict a somewhat larger
$\alpha $ (about $0.53$ and $0.33$ for $k_{\rm B} = 0.003$ and $k_{\rm B} = 0.006$, respectively) 
but this may reflect the simplicity of the model used or other 
complications. Thus the experimental results in Ref. \cite{two}
for the same system but with different cross-link densities 
gave a similar critical velocity $v_{\rm c}$ as observed in Fig. \ref{1logv.2logG.overG0.eps},
but larger $\alpha$ exponents ($0.34$ and $0.5$ for 1:30 and 1:20 cross-linker/base; for the 1:10 cross-linker/base 
too few data points was measured to
determine an accurate $\alpha$ but the obtained results 
indicate an exponent even larger than for the 1:20 case). 
The reason for the spread in the experimentally obtained values for the exponent $\alpha$ is not clear.

\vskip 0.3cm
{\bf 7 Dependency of the work of adhesion on the separation speed and the temperature}

The breaking of the adhesive bond between two solids usually occur by interfacial crack propagation.
However, for very small contact regions 
the bond may break uniformly (simultaneously) over the contact region\cite{small}.
In any case the work of adhesion $w$ is usually defined as the work to separate two flat solid surfaces with the
surfaces parallel. In this section we will consider the separation of two parallel surfaces where the separation
speed is constant. Note that this differ from interfacial crack propagation where
the separation velocity depends on the distance from the crack tip. 

Consider the simplest case of separating the surfaces uniformly at the speed $v$
from the equilibrium position.
In this case $u_{\rm B}(t) = v_z t$ where we have used that the total energy is minimal for $u_{\rm B}=0$.
Using (21) the external force
$$F(t)= K v_z t$$
The force to separate the surfaces per unit surface area is
$F(t) P(t)/b^2$
and the work per unit surface area $w$ to separate the surfaces
$$w= {1\over b^2} \int_0^{t_0} dt \ v_z F(t) P(t)$$
where $v_z t_0 = u_{\rm Bc}$. 
If we introduce $\xi=v_z t/u_{\rm Bc}$ we get
$$w= {K u_{\rm Bc}^2\over b^2} \int_0^1 d\xi \  \xi P(\xi)$$
In the limiting case of zero temperature $P(t)=1$ for $t<t_0$
or $\xi < 1$
so that the work of adhesion $w_1$ at zero temperature
$$w_1 = {K u_{\rm Bc}^2\over b^2} \int_0^1 d\xi \  \xi = { K u_{\rm Bc}^2\over 2 b^2}$$
which also equals $u_{\rm Bc} F_{\rm c}/ 2b^2$. Hence
$${w\over w_1} =  2 \int_0^1 d\xi \  \xi P(\xi)\eqno(41)$$
Using $\xi =v_z t/u_{\rm Bc}$ we also get
$$P(\xi) = P(0) {\rm exp}\left ({-{u_{\rm Bc} \over v_z} \int_0^{\xi} d\xi' \kappa (\xi')}\right )$$
$$ + {u_{\rm Bc}\over v_z} \int_0^{\xi} d\xi' \ \kappa_1(\xi') 
{\rm exp} \left ({-{u_{\rm Bc} \over v_z} \int_{\xi'}^{\xi} d\xi'' \kappa(\xi'')}\right )\eqno(42)$$ 
The rate coefficients $\kappa_1 (\xi)$ and $\kappa (\xi)$ are functions of $u_{\rm B}$  but can be considered
as functions of $\xi$ by replacing $u_{\rm B} = \xi u_{\rm Bc}$. 

One interesting limit of this equation is when 
$v \rightarrow 0$ where (42) reduces to $P=P_0 = \kappa_1(\xi)/\kappa(\xi)$.
This limit follows also directly from the rate equations (22) when $dP/dt=0$ and correspond to 
the assumption that thermal equilibrium occur at all stages in the pull-off. Using $P=P_0$ in (41) gives
$w \approx U_0/b^2$ i.e. the adiabatic work of adhesion. 

For numerical calculations the $\xi''$-integral in (42) is problematic as $v\rightarrow 0$
where only a very small $\xi-\xi'$ interval will effectively contribute. This problem is solved by using
as integration variable instead of $\xi''$ the variable $\eta$ defined by $\xi''= \xi e^{-\eta}$ giving
$d\xi'' = - \xi'' d\eta$ and
$$\int_{\xi'}^{\xi} d\xi''  \ \kappa(\xi'')  = \int_0^{\eta_0} d\eta \ \xi'' \kappa(\xi'') $$ 
where $\eta_0 = {\rm log} (\xi/\xi')$.

\vskip 0.3cm
{\bf Numerical results and discussion}

\begin{figure}[tbp]
\includegraphics[width=0.47\textwidth,angle=0]{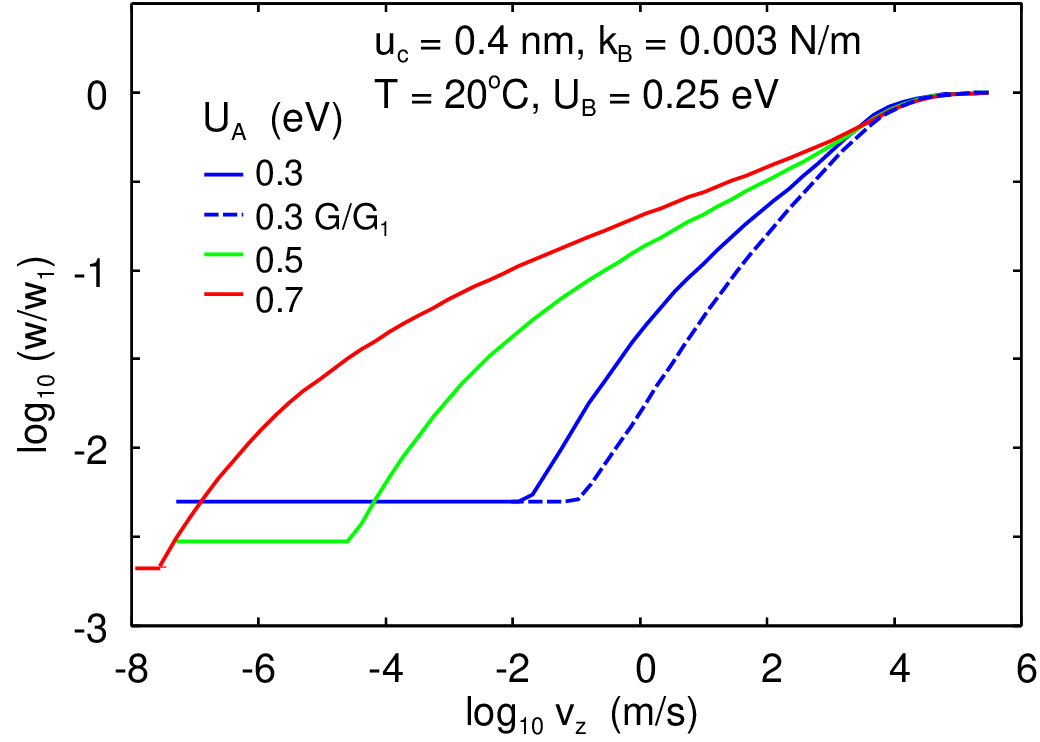}
\caption{
The work of adhesion $w$ in units of its zero temperature value $w_1 = w(T=0)$ as a
function of the pull-off speed (Log-Log scale) 
for the bond energies 
$U_{\rm A}=0.3$, $0.5$ and $0.7  \ {\rm eV}$.
For the temperature $T=20^\circ {\rm C}$ and the spring
constant $k_{\rm B}=0.003 \ {\rm N/m}$, the separation to break the bond $u_{\rm c} = 0.4 \ {\rm nm}$,
and the bond energy $U_{\rm B}=0.25 \ {\rm eV}$. 
For the three cases with $U_{\rm A}=0.3$, $0.5$ and $0.7 \ {\rm eV}$ we have
$\mathcal{E}_1 = w_1 b^2 = 60.4$, $167.4$ and $327.8 \ {\rm eV}$.
}
\label{1logv.2logG0.adh.EA.eps}
\end{figure}

\begin{figure}[tbp]
\includegraphics[width=0.47\textwidth,angle=0]{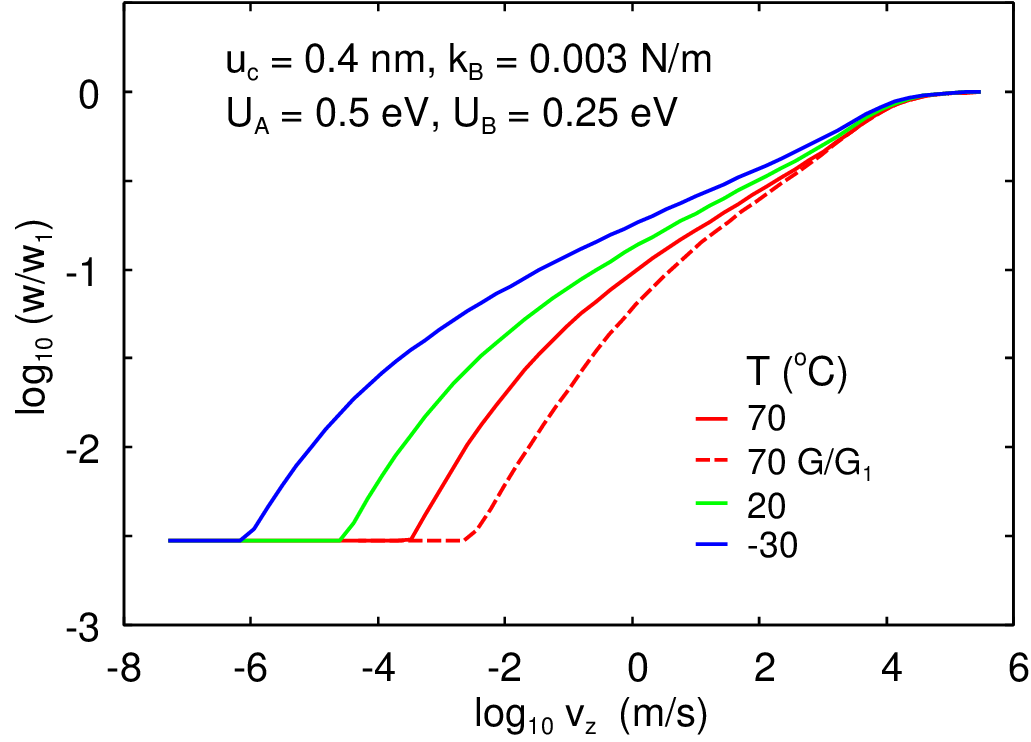}
\caption{
The work of adhesion $w$ in units of its zero temperature value $w_1=w(T=0)$
as a function of the pull-off speed (Log-Log scale) for the temperatures
$T=-30$, $20$ and $70^\circ {\rm C}$
The spring constants
$k_{\rm B} = 0.003$. For the 
A-well bond energy $U_{\rm A}=0.5 \ {\rm eV}$, the separation to break the bond $u_{\rm c} = 0.4 \ {\rm nm}$,
and the B-well bond energy $U_{\rm B}=0.25 \ {\rm eV}$. 
}
\label{1logv.2logG0.temp.adh.eps}
\end{figure}

\begin{figure}[tbp]
\includegraphics[width=0.47\textwidth,angle=0]{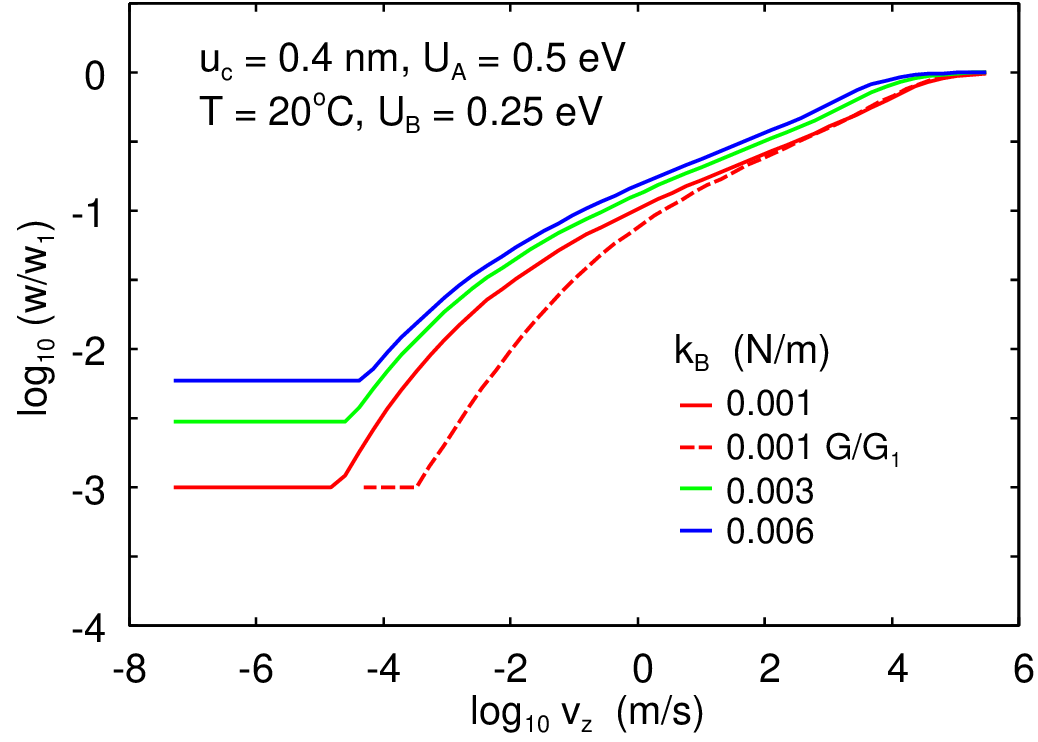}
\caption{
The work of adhesion $w$ in units of its zero temperature value $w_1=w(T=0)$
as a function of the pull-off speed (Log-Log scale) for the spring constants
$k_{\rm B} = 0.001$, $0.0.003$ and $0.006 \ {\rm N/m}$. For $T=20^\circ {\rm C}$, the 
A-well bond energy $U_{\rm A}=0.5 \ {\rm eV}$, the separation to break the bond $u_{\rm c} = 0.4 \ {\rm nm}$,
and the B-well bond energy $U_{\rm B}=0.25 \ {\rm eV}$. 
For the three cases with $k_{\rm B} = 0.001$, $0.0.003$ and $0.006 \ {\rm N/m}$ we have
$\mathcal{E}_1 = w_1 b^2 = 501.2$, $167.4$ and $84.0 \ {\rm eV}$.
}
\label{1logv.2logG0.k.adh.eps}
\end{figure}

Fig. \ref{1logv.2logG0.adh.EA.eps}, \ref{1logv.2logG0.temp.adh.eps} and \ref{1logv.2logG0.k.adh.eps}
shows the work of adhesion $w$ in units of its zero temperature value $w_1 = w(T=0)$ as a
function of the pull-off speed (Log-Log scale) 
for different bond energies (Fig. \ref{1logv.2logG0.adh.EA.eps}), different temperatures (Fig. \ref{1logv.2logG0.temp.adh.eps})
and different spring constants $k_{\rm B}$ (Fig. \ref{1logv.2logG0.k.adh.eps}).
For $U_{\rm A}=0.3$, $0.5$ and $0.7 \ {\rm eV}$ we have
$\mathcal{E}_1 = w_1 b^2 = 60.4$, $167.4$ and $327.8 \ {\rm eV}$.
The dashed lines in the figures shows the crack-energy $G(v)/G_1$ as a function of the crack-tip
speed (Log-Log scale, from Fig. \ref{1logv.2logG.over.GT=0.eps}-\ref{1logv.2logG.varyT.eps}) 
using the same parameters as used for the same colored solid lines.
The work of adhesion curves show similar dependency on the pull-off speed as the
crack-energy on the crack-tip speed, but are shifted along the velocity axis as expected because
the crack-tip speed (in $x$-direction) is not the same as the crack opening speed (in the $z$-direction)
and the latter depends on time even when the crack-tip speed is constant.

\begin{figure}[tbp]
\includegraphics[width=0.35\textwidth,angle=0]{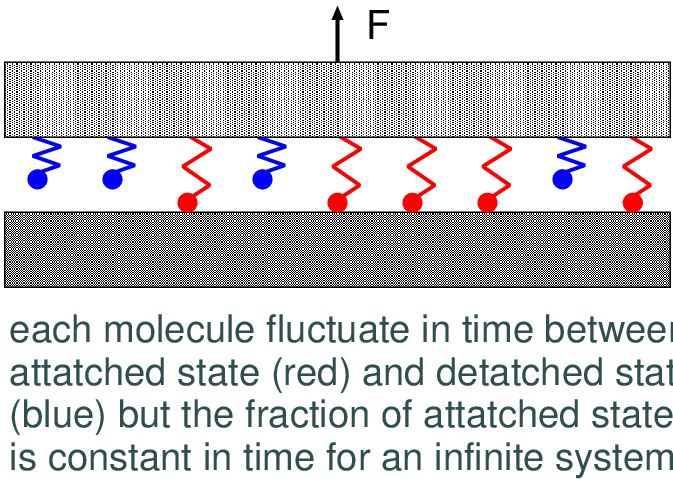}
\caption{
If thermodynamic (kinetic) equilibrium occur any given molecule
will fluctuate (in time) between the bonding and non-bonding states but when considering a large number
of molecules a non-zero fraction of the molecules will be in the bonding state. 
}
\label{FlatFluctuate.eps}
\end{figure}

It is important to note the following fact. For $T>0 \ {\rm K}$ an arbitrary small force can break an
adhesive bond and remove a molecule from a surface. If the bond is strong or the temperature low, 
a very long time will be needed for a large enough thermal fluctuation to occur to break the bond.
Nevertheless, even without an applied force a molecule will desorb from a surface if one wait a long enough time. 
However, for molecules binding two nominally flat and infinite 
large surfaces the situation is different. In this case, even if the bonds are very weak, a small applied
pull-pressure cannot separate the surfaces. The reason is that even if each individual molecule fluctuate between the bonded
and non-bonded state, a non-zero fraction of the molecules will be in the bonded state. That is, 
if thermodynamic (or kinetic) equilibrium occur any given molecule
will fluctuate (in time) between the two states but when considering a large number
of molecules a (constant) fraction of the molecules will be in the bonded state (see Fig. \ref{FlatFluctuate.eps}). 
For this case the work to slowly remove the contact
between the solids will equal to the thermodynamic work of adhesion, which is equal to the change in the
free energy between the system with separated surfaces and the system in contact at the equilibrium separation. 

\vskip 0.3cm
{\bf 8 Cohesive crack propagation}

The stress in the vicinity of cohesive cracks in rubber materials is so high that
linear viscoelasticity is unlikely to be valid even several nm away from the crack tip.
Depending on the material considered one may need to take into account
non-linear elasticity, 
plastic deformation, or the formation of defects, such as cavities or 
stringing, in a region close to the crack tip.   
Most of these processes are strongly temperature dependent and will contribute
via the factor $G_0(v,T)$ to the crack-energy. 

It has recently been suggested\cite{Wang1,Wang2} that the $(v,T)$ dependency
of the dissipation in the crack-tip process zone
may be more important for the crack-energy than the 
viscoelastic factor $[1+f(T,v)]$. This may indeed be the
case for conditions where the viscoelastic effect is very small, 
such as for temperatures much higher than
the rubber glass transition temperature. However, 
in general the viscoelastic factor is extremely important and
increases by a factor of 
$E_{\infty}/E_0$ (where $E_{\infty}$ and $E_0$ are the high and low frequency modulus)
between high and low sliding speeds (or low and high temperatures). 
For ``normal'' rubber compounds $E_{\infty}/E_0 \sim 10^3$, and for weakly crosslinked rubber even more. 

Here we present numerical results illustrating the velocity 
dependency of $G_0(v,T)$ for cohesive cracks, resulting from
stress aided thermally activated bond-breaking. 
We use the same model as in Sec. 6, but the meaning of the $U_{\rm A}$, $U_{\rm B}$, $u_{\rm c}$, $k_{\rm A}$, $k_{\rm B}$
parameters differ. Thus the spring constant $k_{\rm B}$ in Sec. 6 was of entropic origin 
as relevant for breaking of weak adhesive bonds.
For breaking cohesive bonds the chains in the process zone
will stretch so strongly that the entropic elasticity is
irrelevant. In the theory of Lake and Thomas 
it was assumed that just before a chain break at one of the (covalent) bonds, it is stretched
so that all the bonds along the chain are close to the breaking point.
Thus the elastic energy stored a chain consisting of $N^*$ bond-units, 
at the point where a bond will break, 
is $N^* U_{\rm A}$ where $U_{\rm A}$ is the energy to 
break a bond. 
(Note that $N^* >> N$ since the separation between the covalent bonds is $\approx 0.1 \ {\rm nm}$ while
the Kuhn length $b \approx 1 {\rm nm}$.)
In this case the crack-energy (neglecting thermal activation) 
is obtained by multiplying  $N^* U_{\rm A}$ with the number
of chains per unit area crossing the fracture plane. 
Using this approach Lake and Thomas estimated the fracture energy
$G_0$ and found it to be typically of order 
$10 \ {\rm J/m^2}$ or $100 \ {\rm eV/nm^2}$, which is close to what is observed under conditions
where the viscoelastic factor $[1+f(v,T)]$ is small.

\begin{figure}
\centering
\includegraphics[width=0.95\columnwidth]{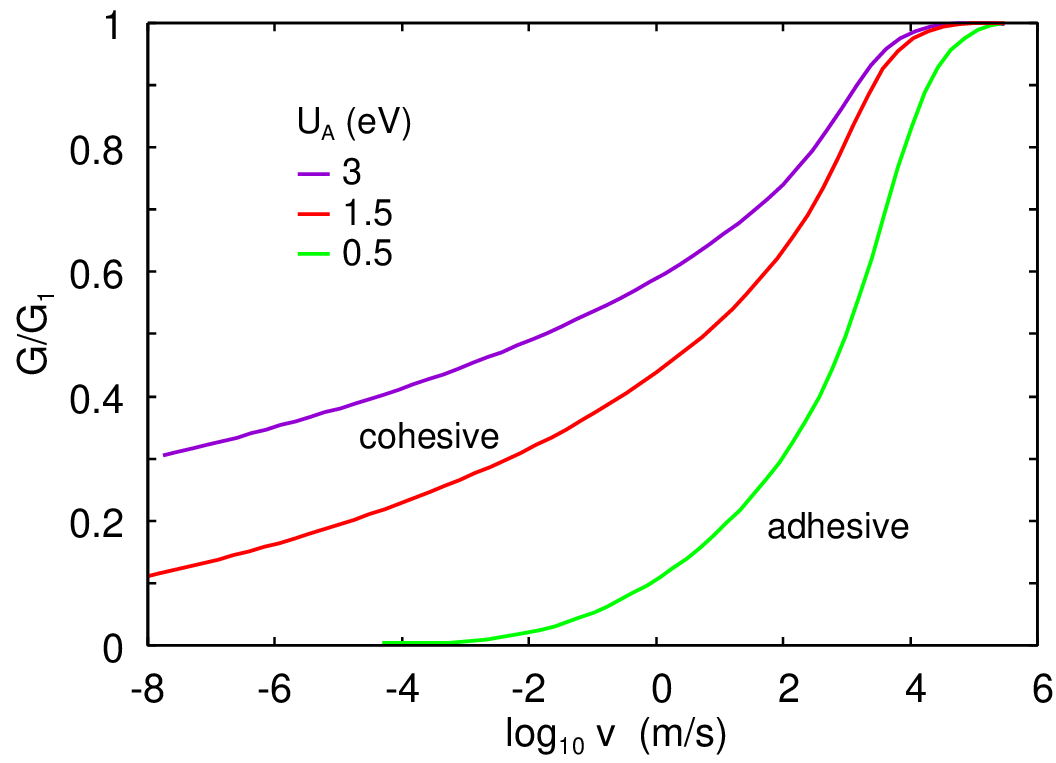}
\caption{\label{1logv.2G0.adhesive.cohesive.eps}
The crack-energy $G$ in units of its zero temperature value $G_1=G(T=0)$
as a function of the crack-tip speed for $T=20^\circ {\rm C}$, 
for conditions typical for adhesive cracks
(red line) and cohesive cracks (blue line). In the calculations for adhesive crack we used
$U_{\rm A} = 0.5 \ {\rm eV}$, $U_{\rm B}=0.25 \ {\rm eV}$, $k_{\rm B}=0.003 \ {\rm N/m}$, 
$k_{\rm A} = 2U_{\rm A}/u_{\rm c}^2 \approx 1.0  \ {\rm N/m}$, $u_{\rm c} = 0.4 \ {\rm nm}$,
and for the cohesive cracks (blue line)
$U_{\rm A} = U_{\rm B}= 3 \ {\rm eV}$, $u_{\rm c} = 0.1 \ {\rm nm}$, 
$k_{\rm A} = 2U_{\rm A}/u_{\rm c}^2 \approx 96.1 \ 
{\rm N/m}$ and $k_{\rm B} = k_{\rm A}/N^*$ 
with $N^*=30$,
and (green line):
$U_{\rm A} = U_{\rm B}= 1.5 \ {\rm eV}$, $u_{\rm c} = 0.1 \ {\rm nm}$, 
$k_{\rm A} = 2U_{\rm A}/u_{\rm c}^2 \approx 48.1 \ {\rm N/m}$ 
and $k_{\rm B} = k_{\rm A}/N^*$ with $N^*=30$.
}
\end{figure}

Fig. \ref{1logv.2G0.adhesive.cohesive.eps} shows the crack-energy $G$ in units of its zero temperature value $G_1=G(T=0)$
as a function of the crack-tip speed for $T=20^\circ {\rm C}$, 
for conditions typical for adhesive cracks
(red line) and cohesive cracks (blue and green lines for $U_{\rm A} = 3$ and $1.5 \ {\rm eV}$, respectively). 
For the cohesive crack (blue line) we have used $U_{\rm A} = U_{\rm B} = 3 \ {\rm eV}$, 
$u_{\rm c} = 0.1 \ {\rm nm}$, $k_{\rm A} = 2 U_{\rm A}/u_{\rm c}^2$
and $k_{\rm B} = k_{\rm A}/N^*$ with $N^*=30$ giving $N^* U_{\rm A} = 90 \ {\rm eV}$. 
Because of the much higher energy needed to break a covalent bond in 
the chain compared to the adhesion bond, the variation of the crack-energy with the 
crack-tip speed is much slower for the cohesive crack. For the adhesive crack the
thermal equilibrium crack-energy is reached already for $v\approx 0.1 \ {\rm mm/s}$
(as also observed in experiments, see Fig. \ref{Combined.ps}) but for the cohesive cracks the thermal
equilibrium value cannot be reached for any crack-tip velocity of human interest.
We conclude that for cohesive cracks the viscoelastic factor $[1+f(v,T)]$ will in general
result in a much stronger velocity dependence than that of the prefactor
$G_0(v,T)$, at least for the bond-breaking process considered here.

\begin{figure}
\centering
\includegraphics[width=0.95\columnwidth]{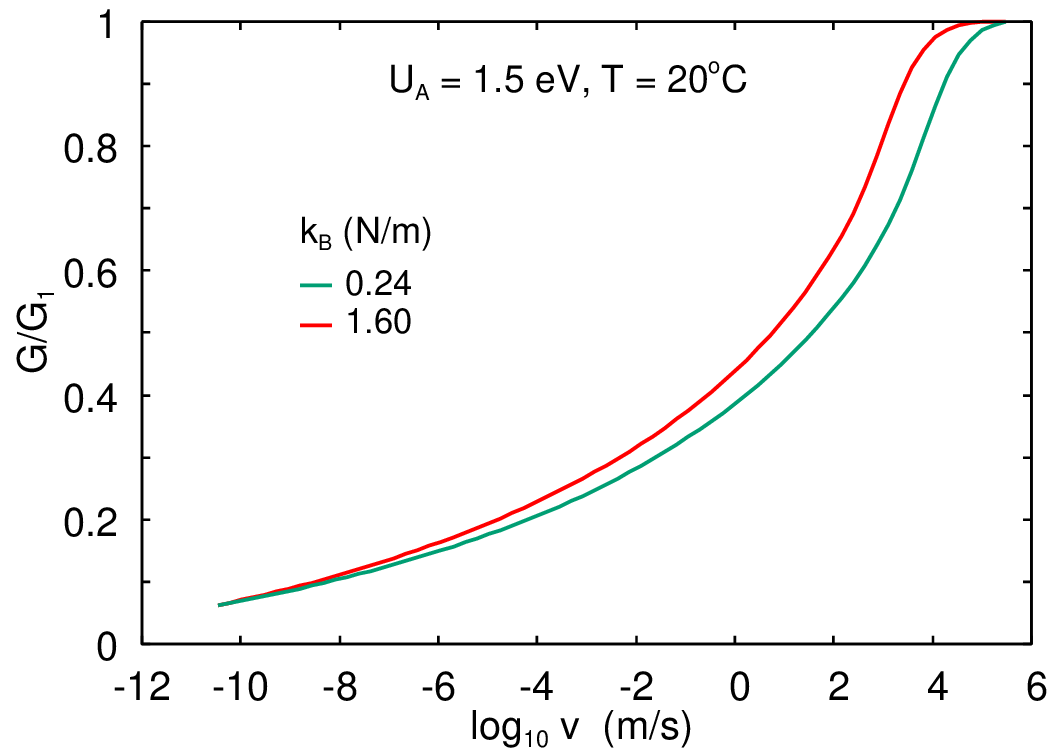}
\caption{\label{1logv.2G.over.G0.UA=1.5eV.N=30.200.eps}
The crack-energy $G$ in units of its zero temperature value $G_1=G(T=0)$
as a function of the crack-tip speed for $T=20^\circ {\rm C}$, 
for conditions typical for cohesive cracks. In the calculations for adhesive crack we used (red line)
$U_{\rm A} = U_{\rm B}= 1.5 \ {\rm eV}$, $u_{\rm c} = 0.1 \ {\rm nm}$, 
$k_{\rm A} = 2U_{\rm A}/u_{\rm c}^2 
\approx 48.1 \ {\rm N/m}$ and $k_{\rm B} = k_{\rm A}/N^*$ with $N^*=30$,
and (green line) for the same parameters except with $N^*=30$ replaced by $N^*=200$.
}
\end{figure}

\begin{figure}
\centering
\includegraphics[width=0.95\columnwidth]{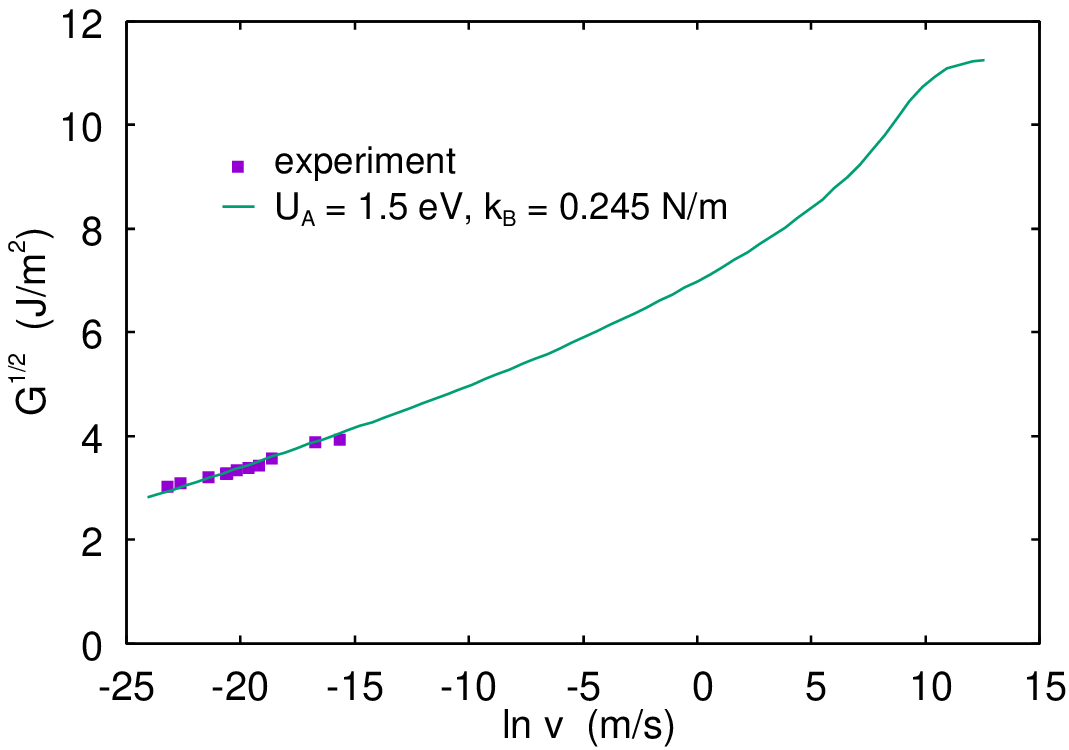}
\caption{\label{1lnv.2sqrtG.theory.experiment.eps}
The square-root $G^{1/2}$ of the crack-energy 
as a function of the (natural) logarithm ${\rm ln} v$ of the crack-tip speed for $T=20^\circ {\rm C}$.
The squares are measured data from Ref. \cite{Chaud2} and the solid line the theory prediction. 
In the calculation we used
$U_{\rm A} = U_{\rm B}= 1.5 \ {\rm eV}$, $u_{\rm c} = 0.1 \ {\rm nm}$, 
$k_{\rm A} = 2U_{\rm A}/u_{\rm c}^2 \approx 48.1 \ {\rm N/m}$ and $k_{\rm B} = 0.245 \ {\rm N/m}$.
}
\end{figure}

Fig. \ref{1logv.2G.over.G0.UA=1.5eV.N=30.200.eps}
shows the crack-energy $G$ in units of its zero temperature value $G_1=G(T=0)$
as a function of the logarithm of the crack-tip speed for $U_{\rm A} = 1.5 \ {\rm eV}$ 
for two different spring constants,  
$k_{\rm B} = 1.6$ (red line) and $0.24  \ {\rm N/m}$ (green line). The difference between
the two cases for very small sliding speeds is very small but this is the case only for the 
ratio $G/G_1$ since $G_1 \sim k_{\rm B}^2$ differ by a factor of $(1.6/0.24)^2 \approx 44.4$.

In a beautiful experimental study Chaudhury has measured the velocity dependency of the fracture
energy for PDMS films chemically bound to silanized silica glass surfaces. Using an elegant experimental set-up
of the JKR-type he was able to measure $G(v)$ at extremely 
small crack tip speeds, from $0.1 \ {\rm nm/s}$ to $100 \ {\rm nm/s}$.
For these low crack tip speeds the used silicon rubber can be considered as a perfect elastic material so the
crack energy is determined by the prefactor $G_0(v,T)$ in equation (1).  
Using a simple picture of the bond breaking at the crack tip, assuming that the bond-breaking occur 
with a constant separation velocity, he showed that for low separation velocities $v_z$ the energy $\mathcal{E}$ to break a bond
$\mathcal{E}^{1/2} \sim {\rm ln} v_z$ 
I have not been able to derive this result analytically
using the theory for $G(v)$ presented above, but the numerical results we now present indicate that $G^{1/2} \sim {\rm ln} v$
holds reasonably well for low crack tip speeds.

Fig. \ref{1lnv.2sqrtG.theory.experiment.eps} shows the relation between $G^{1/2}$ and ${\rm ln} v$ 
for the system studied by Chaudhury. 
The squares are the measured data from Ref. \cite{Chaud2} and the solid line the theory prediction. 
In the calculation have used (39) and (40) with $G_1 = c N^*U_{\rm A}$, $U_{\rm A} = 1.5 \ {\rm eV}$ and $N^*=208$, 
corresponding to the spring constant $k_{\rm B} = 0.245 \ {\rm N/m}$. The number density of polymer chains crossing
the fracture plane was estimated by Chaudhury, using the method of Lake and Thomas, and found to 
be $c = 2.5 \times 10^{18} \ {\rm m^{-2}}$ (see also Appendix B). 
This gives $G_1 = G_0(T=0) = 128 \ {\rm J/m^2}$. 
Note the beautiful agreement between the theory and the experiment. 
Chaudhury argued that the energy to break one bond ($U_{\rm A}=1.5 \ {\rm eV}$) 
is due to siloxane bond scission, which is close to that ($1.5-1.8 \ {\rm eV}$) 
obtained from thermal de-polymerization and stress relaxation kinetics of the siloxane polymers.

\begin{figure}
\centering
\includegraphics[width=1.0\columnwidth]{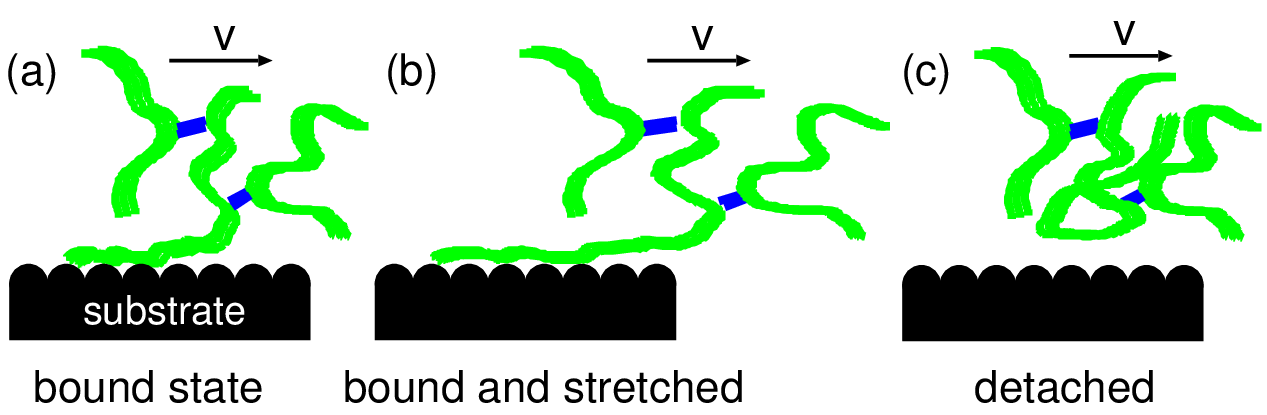}
\caption{\label{SlidingBonds1.eps}
The classical description of a polymer chain at the rubber-block countersurface interface. 
During lateral motion of the rubber block, the chain attach (a), stretches (b), detaches (c), relaxes, and reattaches to the surface to repeat the cycle.
The picture is schematic and in reality no detachment in the vertical direction is expected, 
but only a rearrangement of molecule 
segments (in nanometer-sized domains) parallel to the surface from pinned (commensurate-like) domains to depinned (incommensurate-like) domains.
}
\end{figure}

\vskip 0.3cm
{\bf 9 Discussion}

In the study in Sec. 6 and 7 we have treated $k_{\rm B}$ as an entropic spring
which can be estimated using $k_{\rm B} \approx E b$, where $E$ is the elastic modulus
and $b \sim 1 \ {\rm nm}$ (see the Appendix A). For rubber-like materials this is a good approximation for high 
enough temperature or low enough deformation rates where the
rubber is in the so called ``rubbery'' region.
However, the viscoelastic modulus $E(\omega,T)$ of rubber increases strongly as the deformation frequency $\omega$
increases, and may be $10^3$ (or more) higher in the high frequency ``glassy'' region.
This effect must be taken into account in a more general theory for $G_0 (v,T)$ even if developed on the same simple level 
as in the present study. This generalization can be done in a similar way as described in Ref. \cite{theo2} for the frictional
shear stress (see below).

The picture underlying the theory for adhesive crack propagation presented above is similar 
to the picture of the adhesive contribution to sliding friction proposed by Schallamach\cite{theo0} and illustrated
in Fig. \ref{SlidingBonds1.eps}.  
During lateral motion of the rubber block, the chain attach (a), stretches (b), detaches (c), relaxes, and reattaches to the surface to repeat the cycle.
In reality no ``full'' detachment in the vertical direction is expected, 
but only a rearrangement of molecule segments (in nanometer-sized domains) parallel to the surface 
from pinned (commensurate-like) domains to depinned (incommensurate-like) domains.
The work of Schallamach was extended to more realistic conditions 
by Cherniak and Leonov\cite{theo1} and by Persson and Volokitin\cite{theo2}. In Ref. \cite{theo2} the full
frequency-dependent modulus $E(\omega,T)$ was used when calculating the frictional shear stress.

Support for the importance of the bonding-stretching-debonding contribution to sliding friction was presented
in Ref. \cite{rf1,rf2} (see also Ref. \cite{exp1}) where it was showed that in addition to the viscoelastic
deformations of the rubber surface by the road asperities,
there is an important contribution to the rubber friction from
shear processes in the area of contact. The analysis showed
that the latter contribution could result from bonding-stretching-debonding 
cycles of the type indicated in Fig. \ref{SlidingBonds1.eps}. 
However, the temperature dependency of the shear stress $\tau_{\rm f}$, for temperatures above the rubber glass transition 
temperature $T_{\rm g}$, was found to be weaker than that of the bulk viscoelastic modulus. 
The physical origin of this may relate to the rubber molecule segment mobility at the sliding interface, 
which for rough surfaces may be higher than in the bulk because of increased free-volume effect 
due to the short-wavelength surface roughness. This is consistent with the often observed reduction in the glass transition 
temperature in nanometer-thick surface layers of glassy polymers. 
In Ref. \cite{pullout,Per1} we argued that the opening crack propagation at the asperity contact regions, on very rough surfaces such
as road surfaces, may not give an important contribution to the rubber friction in most cases.

\vskip 0.3cm
{\bf 10 Summary and conclusion}

In this paper I have studied the influence of temperature and the crack-tip velocity of the bond breaking
at the crack tip in rubber-like materials. The bond breaking is considered as a stress-aided
thermally activated process and result in an effective crack propagation energy which 
increases with decreasing temperature or increasing crack-tip speed.
The theory was developed for adhesive cracks (crack propagation between an elastic solid and a counter surface)
but should be valid also for cohesive cracks if the only loss process is bond breaking at the crack tip as in the Lake and Thomas 
model of the crack tip process zone.

The model calculations show a very abrupt onset of non-adiabatic effects with increasing crack-tip speed.
Thus for $v < v_{\rm c}$, where for a typical case of weak adhesion $v_{\rm c} \sim 0.1 \ {\rm mm/s}$, the bond-breaking
occur adiabatically, where the adhesion molecules fluctuate stochastically between the bonded and non-bonded state,
as in thermal equilibrium (liquid-like) dynamics. For $v> v_{\rm c}$ the crack-energy increases rapidly and may be
$\sim 10^3$ bigger at high crack-tip speed than in the adiabatic limit. These results are in reasonable agreement
with experimental data for silicone rubber bound to silica glass surfaces.

For cohesive cracks the velocity dependency of $G_0(v,T)$ for $v>v_{\rm c}$ is much weaker than for
adhesive crack propagation, and the adiabatic limit ($v<v_{\rm c}$) can never be reached on time or velocity scales involved in
human activities. Hence for cohesive cracks in rubber-like materials the viscoelastic factor $[1+f(v,T)]$ will almost always dominate the
velocity (and temperature) dependency of the crack-energy unless the experiments involve
temperatures much higher than the rubber glass transition temperature, or so low crack-tip speed that the rubber response
is in the rubbery part of the viscoelastic spectra.

\vskip 0.3cm
{\bf Acknowledgements}

I thank M.K. Chaudhury for useful discussions and for drawing my attention to his beautiful
work in Ref. \cite{Chaud2}. This work was upported by the Strategic Priority Research Program of the 
Chinese Academy of Science, Grant No. XDB0470200 

\begin{figure}
\centering
\includegraphics[width=0.95\columnwidth]{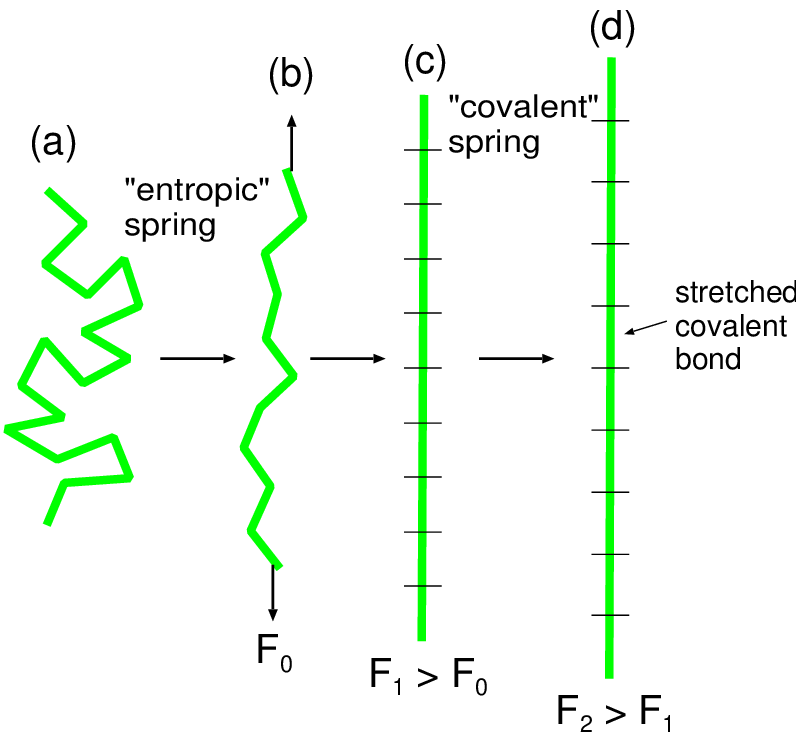}
\caption{\label{EntropicToCovalent.eps}
(a) and (b), adhesive fracture: A chain between a crosslink and a bond to a substrate. 
The chain stretch before the bond to the substrate breaks.
The bond to the substrate is assumed weak and the relevant spring constant is the entropic spring constant.
(c) and (d), cohesive fracture: A chain between two crosslinks on opposite sides of a fracture plane before the fracture. 
When the crosslinks separate at a crack tip the chain stretches before the chain break at any of the bonds along the chain.
In this case the separation force is so high that the relevant spring constant correspond to stretching of the covalent bonds along the chain.
}
\end{figure}

\vskip 0.5cm
{\bf Appendix A: Estimation of the spring constant $k_{\rm B}$}

In this section we denote the spring constant $k_{\rm B}$ by $k$ in order not to mix it up with
the Boltzman constant. There are different ways of estimating the spring 
constant $k$ relevant for adhesive fracture.
If a pressure $p_0$ is applied to a circular area of radius $R$ on an elastic half space 
it will displace the center of the surface with a distance $u = F/k$ where $F=\pi R^2 p_0$
is the applied force and the spring constant 
$$k=\beta R E\eqno(A1)$$ 
where $\beta $ is of order unity (typically 
between $1-2$ depending on the spatial variation of $p_0$, with 
$\beta = \pi /2(1-\nu^2)$ if $p_0$ is constant). Using $R\approx 0.75 \ {\rm nm}$ and
$E\approx 2 \ {\rm MPa}$ (as for fully crosslinked PDMS) we get $k \approx 0.003 \ {\rm N/m}$.

An alternative approach is to use the entropic spring constant [see Fig. \ref{EntropicToCovalent.eps}(a)-(b)] of a chain of length
$L$. Here a ``chain'' denote the part of a parent long-chain molecule lying between adjacent crosslinks 
(see Fig. \ref{breakchain.eps}).
A long-chain molecule act as a spring in tension with the restoring force based on changes in entropy 
rather than energy. This ``entropic-spring'' model has played a central role in the molecular theory of rubber elasticity.
From the simplest ``freely-jointed chain model'' 
$$k= {3 k_{\rm B} T \over L b }\eqno(A2)$$
where $b$ is the length of a rigid (Kuhn) segment (monomer unit) 
and $L=Nb$, where $N$ is the number of Kuhn segments in the chain. 
(Each Kuhn segment can be thought of as if they are freely 
jointed with each other.
Each segment in a freely jointed chain can randomly 
orient in any direction without the influence 
of any forces, independent of the directions taken by other segments.)
For Sylgard 1:10 PDMS the Kuhn length $b\approx 1 \ {\rm nm}$.
Using $N=4$ (or $L= Nb = 4 \ {\rm nm}$ at room temperature 
this gives $k \approx 0.003 \ {\rm N/m}$. 

Using the Young's modulus predicted by the freely-jointed chain model
one can show that the two expressions (A1) and (A2) are essentially equivalent. The Young's modulus
$$ E= {3 k_{\rm B} T \over L d^2}\eqno(A3)$$
where $d\approx 0.5 \ {\rm nm}$ is an effective diameter of the chain.
Using this in (A1) gives
$$k = {3 \beta k_{\rm B} T R \over L d^2}\eqno(A4)$$
which equals (A2) if we choose $\beta R = d^2/b$. Since $b/d\approx 2$ this imply that the
diameter of the circular surface region, where the applied stress 
act in the elastic continuum model in order
to give the same spring constant as obtained from the entropic spring model, is of order the
effective diameter of the chain, which is what one intuitively would expect.

The spring constant estimated above is relevant for breaking weak adhesive bonds.
For breaking cohesive bonds the chains will stretch so strongly that the entropic elasticity is
irrelevant [see Fig. \ref{EntropicToCovalent.eps}(c)-(d)]. 
In the theory of Lake and Thomas it is assumed that a chain to be broken is stretched
so that each (chemical) bond along the chain is close to the point of 
breaking the bond. Thus the energy stored
in the chain with $N^*$ chain bonds at the point where one 
bond break is $N^* U_{\rm A}$ where $U_{\rm A} \approx 3 \ {\rm eV}$ is the bond energy.
If a chain is $L=10 \ {\rm nm}$ and the distance between the bonds $\sim 0.1 \ {\rm nm}$
we get $N^*=100$ as a typical number for the number of chain-bonds.
In this case the crack-energy (neglecting thermal activation) 
is obtained by multiplying  $N^* U_{\rm A}$ with the number
of chains per unit area crossing the fracture plane. 
Using this approach Lake and Thomas estimated the fracture energy
$G_0$ and found it to be typically of 
order $10 \ {\rm J/m^2}$ or $100 \ {\rm eV/nm^2}$ which is close to observed
fracture energies for low crack-tip speeds.

\begin{figure}
\centering
\includegraphics[width=0.7\columnwidth]{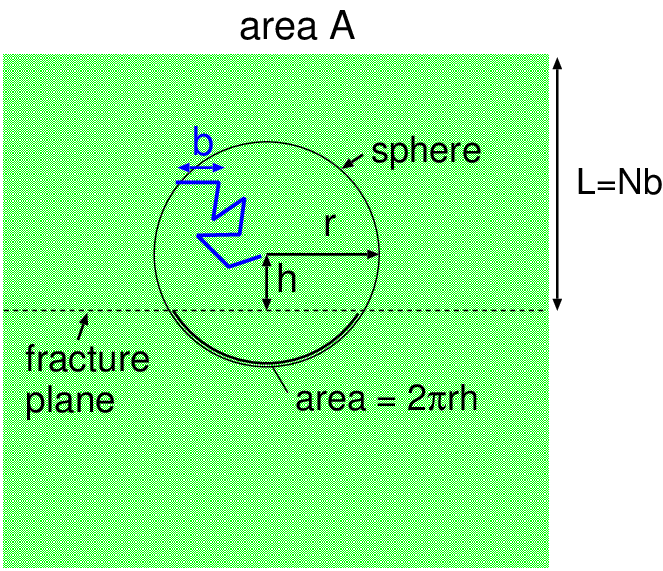}
\caption{\label{SphereCutPlane.eps}
A polymer with the end-to-end separation $r$ and with one end point a distance $h$ from the fracture plane  
has the probability $2\pi r h/ 4\pi r^2 = h/2r$ to cross the fracture plane.
}
\end{figure}

\vskip 0.5cm
{\bf Appendix B: Estimation of density of chain crossing the fracture plane}

The fracture energy of polymers depends on the number of chains per unit area, $c$, crossing the fracture plane.
Here a ``chain'' denote the part of a parent long-chain molecule lying between adjacent crosslinks (see
Fig. \ref{breakchain.eps}).
Lake and Thomas\cite{LT} have estimated $c$ and here we give for the readers convenience a slightly different
derivation.
 
The maximum length of a chain is $L=N b$, where $N$ is the number of Kuhn segments.
Assume that the fracture plane is at $z=0$. 
Any chain in the region $-L < z < L$ has some probability to extend accross the fracture plane.
If $A$ is the surface area the volume of the slab $-L < z < L$ is $V=2AL$. If it containes $n$ chain molecules
then $V= n Ld^2$ where $Ld^2$ is the volume of a chain molecule of length $L$ and effective (chain) diameter $d$.
Using $V=2AL = nLd^2$ the number of chains per unit surface area in the region $-L < z < L$ is $n/A = 2/d^2$ 
Of these chains only a fraction $p$ will cross the fracture plane. Thus the consentration of chains crossing the fracture plane is
$$c = {2p \over d^2}\eqno(B1)$$

For a Gaussian chain the probability that it has the end-to-end length $r$ is
$$P(r) = 4\pi r^2 \left ({\alpha \over \pi} \right )^{3/2} e^{-\alpha r^2}$$
where 
$$\alpha = {3\over 2Nb^2} . \eqno(B2)$$
This probability distribution is normlized so that
$$\int_0^\infty dr \ P(r)=1. $$
Of all chains of length $r$ only those at a distance $h <r$ from the fracture plane has a chance to cross the fracture
plane. A chain of length $r$ can have any angular orientation of the end-to-end vector so the 
pobability that a chain of length $r$ with one endpoint at a distance $h$ from the fracture plane 
will cross the fracture plane is given by the area ratio $A(h)/A_0$, where $A_0 = 4\pi r^2$ and $A(h) = 2 \pi r h$ 
(see Fig. \ref{SphereCutPlane.eps}).
Thus $A(h)/A_0 = h/2r$ and the probability that an arbitrary choosen chain will cross the surface $z=0$ is
$$p = {1\over L} \int_0^L dh \int_0^\infty dr \ P(r) \theta (r-h) {h\over 2r} = {1\over 2 L (\pi \alpha)^{1/2}}$$  
Using (B1) and (B2) this gives
$$c = \left ({2\over 3 \pi}\right )^{1/2} {1\over d^2 \surd N}$$

\end{document}